\documentclass[twocolumn]{aastex6}

\usepackage{ mathrsfs }
\usepackage{graphicx}
\usepackage{amssymb}
\usepackage{amsmath}

\usepackage{longtable}

\def\apjl{ApJ}%
\def\apjs{ApJS}%
\def\ao{Appl.~Opt.}%
\def\aap{A\&A}%
\shorttitle{The MSP mass distribution}
\shortauthors{J. Antoniadis {\it et~al.}}

\begin{document}

\title{The millisecond pulsar mass distribution: \\ 
Evidence for Bimodality and Constraints on the maximum neutron star mass}

\author{John Antoniadis,\altaffilmark{1} Thomas M. Tauris,\altaffilmark{2,3} 
Feryal \"{O}zel,\altaffilmark{4} \\ Ewan Barr,\altaffilmark{5},  David J. Champion, \altaffilmark{2} and  Paulo C. C. Freire\altaffilmark{2}  
}

\altaffiltext{1}{Dunlap Institute for Astronomy \& Astrophysics, University of Toronto, 
50 St. George Street
Toronto, M5S 3H4, Ontario, Canada,  \\ antoniadis@dunlap.utoronto.ca}
\altaffiltext{2}{Max-Planck-Institut f\"{u}r Radioastronomie, Auf dem H\"{u}gel 69, Bonn, 53121, Germany}
\altaffiltext{3}{Argelander Institut f\"{u}r Astronomie, Auf dem H\"{u}gel 71,  Bonn, 53121, Germany}
\altaffiltext{4}{Department of Astronomy, University of Arizona, Tucson, AZ 85721, USA}
\altaffiltext{5}{Centre for Astrophysics \& Supercomputing, Swinburne University of Technology, P.O. Box 218, Hawthorn, Victoria, 3122, Australia}
\

\begin{abstract}
The mass function of neutron stars (NSs) contains information about
the late evolution of massive stars, the supernova explosion mechanism, 
and the equation-of-state of cold, nuclear matter beyond the nuclear
saturation density.  A number of recent NS mass measurements in binary
millisecond pulsar (MSP) systems increase the fraction of massive NSs
(with $M > 1.8$\,M$_{\odot}$) to $\sim 20\% $ of the observed
population.  In light of these results, we employ a Bayesian framework
to revisit the MSP mass distribution.  We find that a single Gaussian model 
 does not sufficiently describe the observed population.  We test alternative empirical
models and infer that the MSP mass distribution is strongly
asymmetric.  The diversity in spin and orbital properties of high-mass
NSs suggests that this is most likely not a result of the recycling
process, but rather reflects differences in the NS birth masses.  The
asymmetry is best accounted for by a bimodal distribution with a low
mass component centred at $1.393_{-0.029}^{+0.031}$\,M$_{\odot}$ and dispersed by $0.064_{-0.025}^{+0.064}$\,M$_{\odot}$, and a high-mass
component with a mean of $1.807_{-0.132}^{+0.081}$ and a dispersion of
$0.177_{-0.072}^{+0.115}$\,M$_{\odot}$.  We also establish a lower
limit of $M_{\rm max}\ge 2.018$\,M$_{\odot}$ at 98\% C.L. for the
maximum NS mass, from the absence of a high-mass truncation in the
observed masses.   Using our inferred model, we find that the measurement of
350 MSP masses, expected after the conclusion of pulsar surveys with
the Square-Kilometre Array, can result in a precise localization of a
maximum mass up to 2.15\,M$_{\odot}$, with a 5\% accuracy. 
Finally, we identify possible massive NSs within the known pulsar population and discuss birth masses of MSPs.
 \end{abstract}


\section{Introduction}
Neutron star (NS) mass measurements are motivated by central questions
in physics and astrophysics, such as the final stages of stellar
nucleosynthesis and mass loss, the supernova (SN) explosion mechanism,
the properties of nuclear interactions, and the gravitational
interaction in strong-field conditions.

At the most fundamental level the structure of NSs is determined by gravity and nuclear interactions. 
Below a critical threshold around $0.1-0.3$\,M$_{\odot}$, 
neutron decay likely leads to rapid decompression and, ultimately, explosion of the star \citep{cst89,hzd02}. 
For larger masses, the relativistic structure
equations \citep{tol39,ov39}\footnote{In what follows, we assume that
  General Relativity holds in the NS interior.} coupled with a model
for microscopic interactions (represented with the Equation of State,
EoS), define a mass-radius (M-R) relation typically characterized by a
canonical radius as well as a maximum mass above which NSs collapse to
black holes (BHs). While the EoS and corresponding M-R relation may be
directly derived from first-principle quantum chromodynamics
calculations, practical limitations due to the difficulty of capturing
the many-body interactions at play necessitate approximations.  In the
absence of experimental data, these calculations lead to a diverse
range of predictions.

Owing to the properties of the M-R relation, simultaneous measurements
of masses and radii, as well as observations of high-mass NSs have the
potential to place stringent limits on the EoS. NS radius measurements
are recently becoming constraining, with $\approx 15$ carried out to
date \citep{of16}. Systematic and modeling uncertainties have been
addressed in numerous studies but some still need to be resolved 
\citep[see e.g.,][and references therein]{gop12,hcl+14,opg+15,nsk+15}.
Similarly, the recent measurements of two extremely high-mass NSs \citep[with
  $M_{\rm NS}\sim 2.0$\,M$_{\odot}$][]{dpr+10,afw+13} place
stringent limits on the EoS (Section\,5) at ultra-high densities, but still leave a wide range
of possibilities \citep[e.g.,][]{lp01,of16}.

While additional mass and radius measurements could help resolve those
remaining uncertainties, numerous questions also remain in the evolution of
massive stars and supernova (SN) explosion mechanism that can be addressed
by studying the NS mass function. NS masses are not expected to be
uniformly distributed between the theoretical extrema, but rather to
cluster around a small number of characteristic values.  In the
textbook example of NS formation, the core of a massive star collapses
when it surpasses the Chandrasekhar limit,
\begin{equation}
M_{\rm {ch}} \simeq 5.83Y_{\rm {e}}^{2}\,{\rm M}_{\odot}.
\end{equation}
Typical iron cores have average electron fractions of $Y_{\rm e}\simeq
0.45$ yielding $M_{\rm {ch}}\simeq 1.18$\,M$_{\odot}$.  In practice,
one needs to apply several corrections, e.g., taking into account the
core's thermal structure, finite entropy, electrostatic interactions
and surface boundary pressure, non-radial convective effects as well
as neutrino radiation during the SN. All of these place the lower end
of the proto-NS gravitational mass between 1.1 and
1.3\,M$_{\odot}$ \citep[see, e.g.,][]{ttw96}.  Further uncertainties
arise from the explosion energy and the location of the mass cut
during the SN, as well as the final stages of nuclear shell burning
\citep{ww95,whw02,lan12}.  In addition, the final remnant may gain
significant mass due to fall-back of material from the stellar
envelope \citep{fw02}.

State-of-the art numerical simulations and 
analytic calculations for core-collapse SNe and their
progenitors predict NS initial mass functions ranging from uni-modal
to highly skewed and/or multi-modal distributions
\citep{ttw96,ujma12,jan12,pt15,ejw+15,sew15,mhlc16}.  Akin to this work, some
studies find notable differences between remnants originating from
stars that burn carbon radiatively or convectively
\citep{ttw96,bhl+01}.  This bifurcation may lead to a bimodal NS mass
distribution. Furthermore, recent studies of core-collapse progenitors consistently find a highly non-linear relation between the initial (or core helium) stellar mass and the final remnant mass \citep[e.g.][ and references therein]{mhlc16}. 

 Additional components may arise due to alternative
formation channels, such as an electron-capture implosion, which is expected to
produce a distinct peak with a small dispersion around
1.25\,M$_{\odot}$ \citep{nom87,plp+04}.  Finally, two significant but
still poorly understood factors for the outcome of massive star
evolution, besides the initial mass, are the effects of wind mass loss, 
and the dynamical interaction and mass transfer in a binary
system \citep{wl99,bhl+01,plp+04}.

Following birth, the NS mass can further increase due to matter
accretion from a binary companion \citep{bvh91,tv06}.  Depending on
the rate and duration of mass transfer, a significant amount of
material may be accreted onto the NS, potentially even driving the
star beyond the critical limit for collapse into a BH.  On average,
larger masses (typically $\ga 0.1$\,M$_{\odot}$) are expected for
``recycled'' millisecond pulsars (MSPs) with low-mass companions, that
have undergone a long episode of stable mass transfer
\citep{tv06,tlk12}

As of today, NS masses have been inferred for $\sim 75$ NSs in X-ray
binaries, double NS systems (DNS) and MSPs \citep{of16}. If one excludes marginal
measurements and strongly model-dependent or probabilistic inferences,
then the sample of reliable, precision measurements reduces to 32
(Fig.\,1) among the DNS and MSP populations.  Notably, all of these
are at least partly based on the radio timing technique, a summary of
which is given in Section\,2.
 
 \begin{figure}
\includegraphics[width=0.5\textwidth]{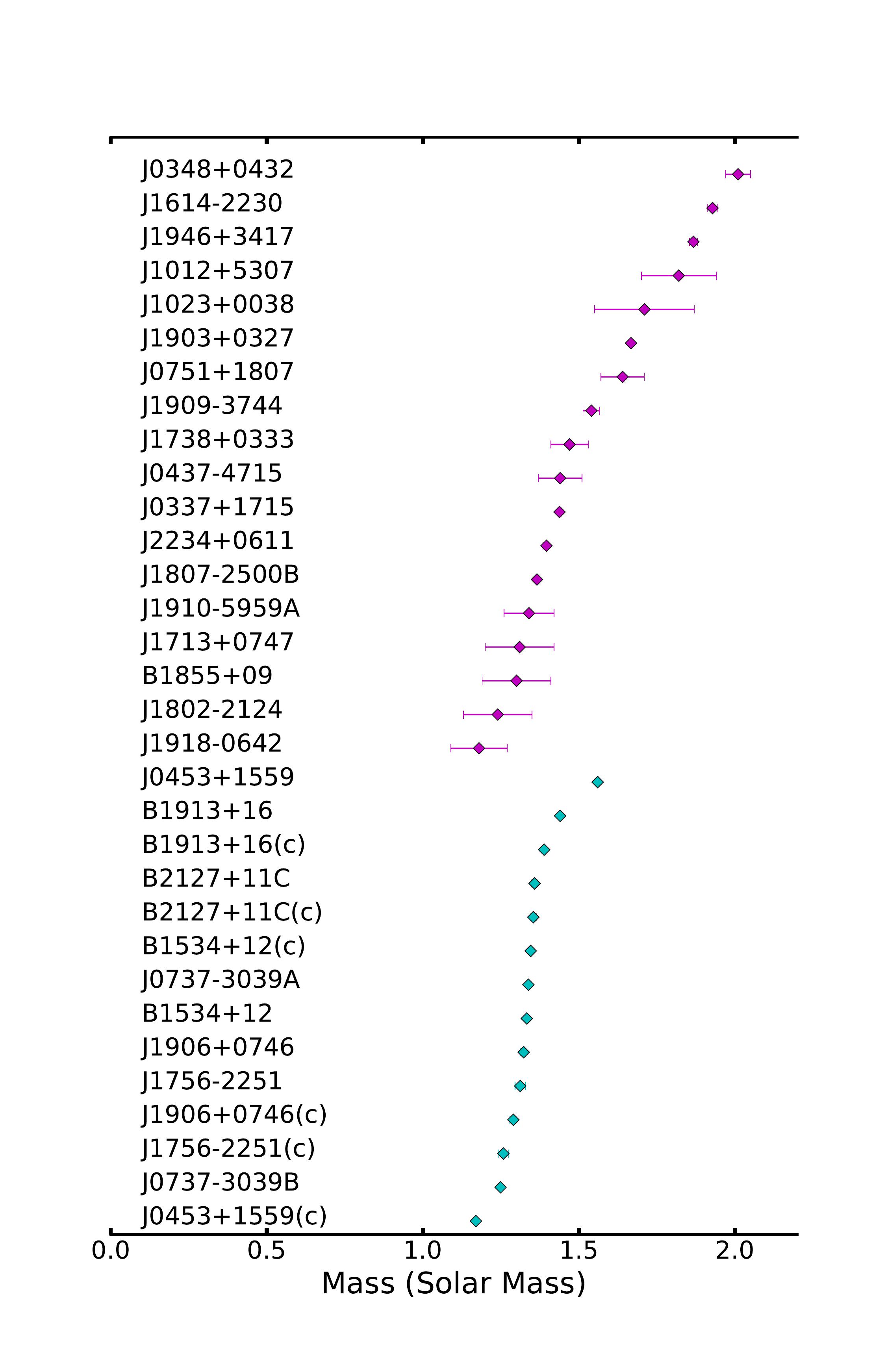}
\caption{Mass measurements and 68\% uncertainty intervals for NSs in
  DNS (blue) or MSP (purple) systems. See \cite{of16} and references therein for the masses of DNS systems.}
\label{fig:1}
\end{figure}

Past attempts to infer the underlying mass distribution based on
growing subsets of these data suggest a strong clustering of masses
between $\sim1.3$ and 1.5\,M$_{\odot}$
\citep{finn94,tc99,spr10,zwz+11,vrh11,opn+12,kkd+13}.  Recent studies
by \cite{opn+12} and \cite{kkd+13} distinguish between different NS
types and find statistically significant differences between those
believed to be close to their birth masses and the ones that have
undergone at least one long-term accretion episode.  A distinct
property of the former NS type, as manifested in the DNS mass
distribution, is a relatively small dispersion of only $\Delta M
\simeq 0.05$\,M$_{\odot}$ around the mean mass of $M \simeq
1.35$\,M$_{\odot}$ \citep{opn+12}. As argued by \cite{opn+12}, it
is possible that the small dispersion reflects a highly tuned
formation channel for DNSs. For example, this likely implies
inefficient but precise amount of accretion of fall-back material
during the SN, which may be difficult to understand in the context of
stellar evolution \citep{jan12,wjm13}.

Recent developments in pulsar searches and timing have led to a nearly
exponential increase in the mass measurements of MSPs and, in
particular, in the discovery of some fairly massive pulsars
\citep{of16}. Mass distributions inferred based on earlier data do not
predict many massive ones. For example, \cite{opn+12} expect about $5-7$\%
of MSPs with masses above $1.8$\,M$_{\odot}$, whereas the new
discoveries of J0348+0432 \citep[$M=2.01(4)$\,M$_{\odot}$;] []{afw+13}
and J1946+3417 \citep[$M=1.867(13)$\,M$_{\odot}$][]{barr16}, as well
as a number of other mass refinements (see Fig.\,1) suggest the
actual fraction to be larger than 20\%.  These systems have very
distinct orbital properties, and their masses have been measured with
different methods. It is therefore unlikely that the new masses result
from selection effects caused by observational bias.

In this paper, we model the MSP mass distribution using the most
up-to-date ensemble of mass measurements. We compare uni-modal and
bi-modal approximations using Bayesian inference techniques and find
that the bimodal distribution in the MSP masses is preferred by the
current data. We examine the implications of these different intrinsic
distributions for stellar evolution and the EoS. Furthermore, we use
our findings to make zero-order estimates for future large-scale
pulsar surveys, such as those planned for the Square Kilometre Array
(SKA). The lay-out of the paper is as follows: In Section \,2 we
provide a brief overview of mass measurement methods and discuss our
dataset. In Section\,3 we outline our statistical method and then
present our main results in Section\,4.  We examine the ramifications
for the EoS in Section\,5.  Finally, we conclude with a broader
discussion in Section\,6.

\section{Millisecond Pulsar masses} 

Pulsar mass measurements can be obtained using a broad range of
techniques at different wavelengths. For MSPs, most constraints come
from precision radio timing, sometimes supplemented by optical
observations of their binary companions. In what follows, we shall
briefly review these methods and discuss their strengths and
weaknesses.

\subsection{Radio Timing}
Radio timing observations of binary pulsars yield precise measurements
of the orbital period $P_{\rm b}$ and projected semi-major axis,
$x\equiv a_{p}\sin i$.  These quantities allow to determine the mass
function,

\begin{equation}
f(m_{\rm p},m_{\rm c},i) = \frac{(m_{\rm c}\sin i)^3}{ (m_{\rm p} + m_{\rm c})^2} 
= \left(\frac{2\pi} {P_{\rm b}}\right)^2 \frac{x^3}{G},
\end{equation}
which relates the unknown stellar masses, 
$m_{\rm p}$ and $m_{\rm c}$, and inclination, $i$. 

Because Eq.\,2 connects three unknowns, inference of the pulsar mass
requires the measurement of at least two additional quantities that
depend on those parameters.  For sufficiently compact binaries, this
can be achieved with the measurement of post-Keplerian (pK) parameters
induced by relativistic effects.  These include the precession of the
orbital periastron $\dot{\omega}$, the Einstein-delay $\gamma$ (which
accounts for time-dilation effects and the varying gravitational
redshift along the orbit), the Shapiro-delay $\Delta t_{\rm s}$, as
modelled by the parameters $r$ and $s$ (describing the extra
travel-time due to the companion's gravitational potential), and the
orbital decay $\dot{P}^{\rm GW}_{\rm b}$ due to emission of
gravitational waves.  In General Relativity (GR) the pK parameters
become functions of the stellar masses and Keplerian parameters
\citep[see ][for details]{lk05}:

\begin{equation}
\dot{\omega} = 3 \left( \frac{P_{\rm b}}{2\pi} \right)^{-5/3} \left( T_{\odot}M_{\rm T} 
\right)^{2/3} \left(1-e^2 \right)^{-1},
\end{equation}
\begin{equation}
\gamma = e \left( \frac{P_{\rm b}}{2\pi} \right)^{1/3} T^{2/3}_{\odot} M_{\rm T}^{-4/3} 
m_{\rm c} \left(m_{\rm p}+2m_{\rm c} \right),
\end{equation}
\begin{equation}
r=T_{\odot}m_{\rm c},
\end{equation}
\begin{equation}
s=\sin i = x \left(\frac{P_{\rm b}}{2\pi}\right)^{-2/3} T_{\odot}^{-1/3} M_{\rm T}^{2/3}m^{-1}_{\rm c},
\end{equation}
\begin{multline}
\dot{P}^{\rm GW}_{\rm b} = -\frac{192\pi}{5}  \left( 1 + \frac{73}{24}e^2 +  \frac{37}{96}e^4 \right) \times \\  \times
\left(1-e^2 \right)^{-7/2} \left(\frac{2\pi \mathcal{M} T_{\odot}}{P_{\rm b}}\right)^{5/3},
\end{multline}
where $T_{\odot} \equiv G M_{\odot} / c^3 = 4.925490947$\,$\mu$s is
the solar mass in time units, $M_{\rm T}=m_{\rm c}+m_{\rm p}$ is the
total mass of the binary, and $\mathcal{M} = (m_{\rm p} m_{\rm
  c})^{3/5} (m_{\rm p} + m_{\rm c})^{-1/5}$ is the chirp mass of the
system.

Due to their formation history, most binary MSPs in the Galactic disk
have eccentricities of order $10^{-7} - 10^{-3}$, rendering the
measurement of $\dot{\omega}$ and $\gamma$ extremely challenging.
Similarly, the Shapiro delay magnitude depends sensitively on the
inclination, and is typically relevant only for systems viewed nearly
edge-on.  Finally, the measurement of $\dot{P}^{\rm GW}_{\rm b}$ is
only possible in extremely compact binary MSPs (P$_{\rm b}
\lesssim 1$\,d) with point-mass like companions (i.e., in double NS and
NS--white dwarf binaries).

On the other hand, a substantial number of MSPs in globular clusters,
as well as a handful of systems in the Galactic field have
sufficiently high eccentricities to allow for constraints on
$\dot{\omega}$ and consequently the total mass $M_{\rm T}$
\citep[see][and references therein]{ant14,vf14}.

\subsection{Optical Spectroscopy} 
Additional information on the masses can also be obtained when the
pulsar companion has an optically bright counterpart. Phase-resolved
spectroscopy yields the orbital radial velocity amplitude $K_{\rm c}$,
which together with $x$ and $P_{\rm b}$ for the pulsar, yields the
mass ratio of the binary, $q \equiv m_{\rm p}/ m_{\rm c} = K_{\rm
  c}/K_{\rm p}$.  Furthermore, the spectrum of the companion contains 
information about its composition and atmospheric properties, which in
turn depend on the stellar mass and radius.

Most known MSPs have He-core white-dwarf companions with a pure
hydrogen atmosphere. Despite the model dependences implicit in the
spectroscopic method, the mapping between atmospheric parameters and
white-dwarf masses has reached a sufficient level of precision to
allow for accurate mass determinations \citep[see][and references
  therein]{antoniadis, itla14, tls+13, amc13, tgk+15}.

\subsection{MSP mass measurements and uncertainties} 
The sample of MSPs collected here consists of systems with constraints
on at least the total mass $M_{\rm T}$, or mass ratio $q$.  Compared
to previous work, our definition of MSPs slightly differs. Instead of
selecting our sample solely based on the pulsar spin period, we make
choices on a case-by-case basis, taking into consideration other
observed properties such as the orbital period and companion type.
For example, the massive pulsar PSR\,J0348+0432 with $P_{\rm s} =
39$\,ms, would not normally qualify as an MSP ($P_{\rm s} \lesssim
30$\,ms).  Nevertheless, the system most likely evolved from a
low-mass X-ray binary (LMXB) and therefore might have experienced
significant accretion (recycling) from its binary companion
\citep{antoniadis,itl14,itla14}.

Overall, our sample consists of $19$ MSPs with precisely determined
masses, 10 MSPs with constraints only on $M_{\rm T}$, and 4 systems
with constraints on $q$. We show the likelihoods over mass for each of
these pulsars in Fig.\,2 and describe them in more detail below.

 \begin{figure}
\includegraphics[width=0.5\textwidth]{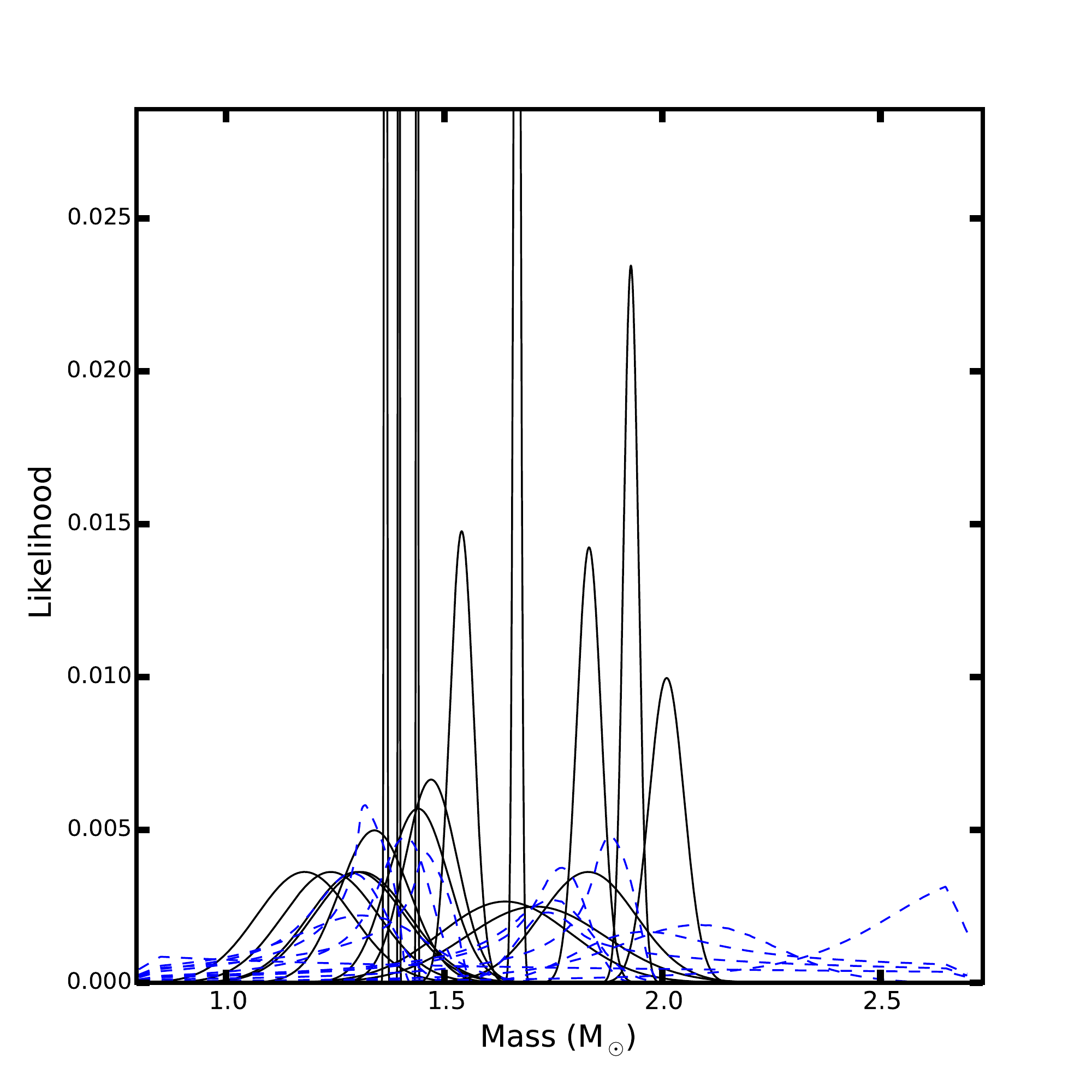}
\caption{Mass likelihoods for the systems in Tables 1--3, based on
  Eq.\,1 (Table\,1, solid lines) and Eqs.\,9 \& 10 (dashed lines). All
  likelihoods are normalized so that the enclosed area is the
  same. The three precise MSP measurements in the 
  $1.3-1.4$\,M$_{\odot}$ range, have peak likelihoods around $\sim 0.35$.}
\label{fig:2}
\end{figure}

\subsubsection{Precision mass measurements}
These systems are shown in Table\,1.  The sub-sample primarily
consists of pulsars with constraints on either two pK parameters, or
spectroscopically resolved He~white dwarf companions.  We also include
PSR\,J0337+1715 \citep{rsa+14}, a pulsar in a hierarchical triple
system, where the masses are obtained from the timing signature of the
3-body interactions, and PSR\,J1023+0038 \citep{asr+09}, a
transitional MSP with a measured parallax and an optically-bright
companion.  For these binaries, we write the likelihood that pulsar
$j$ has a mass $m_{\rm p}$ as:

\begin{equation}
E_{j}({\rm data}|m_{\rm p}) = C_{j}\exp\left[-\frac{(m_{\rm p} - m_0^j)^2}{2\sigma_{m_0^j}
^2}\right],
\end{equation}
where $\sigma_{m_0^j}$ is the inferred measurement uncertainty on
$m_{\rm p}$ and $C_{j}$ a proper normalization factor to ensure that
$\int E_{j}d{m_{\rm p}} = 1$. The mass likelihoods for the systems in
Table\,1 based on Eq.\,8 are shown in Fig.\,2 as solid lines.

It is worth noting that for systems studied with optical spectroscopy,
the actual estimate is slightly asymmetric around the mean, with a
skewness towards larger masses. However, for the systems considered
here, this asymmetry is small and therefore can be safely accounted
for with an appropriate increase in $\sigma_{m_0^j}$.

\begin{table*}
\begin{center}
\caption{Radio Millisecond pulsars with precise mass measurements}
\begin{tabular}{llrl}

\# &PSR Name & Mass [$M_{\odot}$] & Reference \\ 
\hline 
\hline 
1 &J0337+1715  &	1.4378(13)& \cite{rsa+14}      \\

2 &J0348+0432  &	2.01(4)   & \cite{afw+13}          \\

3 &J0437$-$4715 &	1.44(7)  & \cite{rhc+16}        \\

4 &J0751+1807 &	1.64(15)   & \cite{dcl+16}             \\

5 &J1012+0507 &	1.83(11)  & this work (Appendix)       \\

6 &J1023+0038 &	1.71(16)  & \cite{dab+12}           \\

7 &J1614$-$2230 &	1.928(17) & \cite{fpe+16}   \\

8 &J1713+0747 &	1.31(11)  & \cite{zsd+15}    \\

9 &J1738+0333 &	1.47(7)	  & \cite{avk+12}           \\

10 &J1802$-$2124 &	1.24(11)  & \cite{fsk+10}           \\

11 &J1807$-$2500B &	1.3655(21)& \cite{lfrj12}         \\

12 &B1855+09  &	1.30(11)  & \cite{fpe+16}            \\

13 &J1903+0327 &	1.667(7) & \cite{fbw+11}          \\

14 &J1909$-$3744 &	1.540(27) & \cite{dcl+16}          \\

15 &J1910$-$5959A &	1.34(8)	  & \cite{cbp+12}           \\
16 & J1918$-$0642 & 1.18(11)       & \cite{fpe+16} \\
17 &J1946+3417 &	1.832(13) & \cite{barr16}         \\

18 &J2234+0611 &	1.396(11) & \cite{sto16}        \\



\hline 
\end{tabular} 
\end{center}
\end{table*}

\subsubsection{Pulsars with constraints on the total mass}
This group includes systems with a constraint on the total mass
$M_{\rm T}$ (Table\,2). Assuming an inclination with a probability
distribution that is uniform in $\cos i$, the likelihood
for the mass of the $j^{\rm th}$ pulsar can be written as

\begin{multline}
E_{j}( {\rm data}| m_{\rm p}) = C_j\int dM_{\rm T} \int d(\cos i) \times \\ \times \delta\left[f_{0} - 
f(M_{\rm T},m_{\rm p},i)\right]  \times \exp
\left[ - \frac{(M_{T} - M_{0}^j)^2}{2\sigma^2_{M_{0}^j}}\right],
\end{multline}
where again $C_j$ is a normalization coefficient.  For each $i$, the
Dirac delta function involving the mass function can be evaluated from
$\delta (i - i_{\rm 0})$, where $i_{0}$ is the solution to the mass
function equation for a given set of stellar masses (see
\citealt{opn+12} for details).

\begin{table*}
\begin{center}
\caption{Millisecond pulsar binaries with constraints on the total mass}
\begin{tabular}{llrll}
\# & PSR Name & $f(m)$ [M\,$_{\odot}$]  & M$_{\rm{T}}$ [M\,$_{\odot}$] & Reference \\ 
\hline 
\hline 
1   & J0024$-$7204H & 0.001927	     &  1.61(4) &\cite{fck+03} \\
2   & J0514$-$4002A & 0.14549547     &  2.453(14)& \cite{frg07} \\
3   & J0621+1002       & 0.027026849    &  2.32(8)	 &\cite{sna+02} \\
4   & B1516+02B        & 0.000646723  &  2.29(17) &\cite{fwbh08} \\
5   & J1748$-$2021A & 0.0518649	     &	1.97(15) & \cite{fwbh08} \\
6   & J1748$-$2021B & 0.0002266235 &	2.92(20) &\cite{frb+08} \\
7   & J1748$-$2446I   & 0.003658        &	2.17(2)	&\cite{rhs+05} \\
8   & J1748$-$2446J  & 0.013066	        &	2.20(4)	&\cite{rhs+05} \\
9   & B1802$-$07       & 0.00945034     &	1.62(7)	&\cite{tc99} \\
10 & J1824-2452C     & 0.006553	         &	1.616(7) &\cite{frb+08} \\
\hline 
\end{tabular}
\end{center}
\end{table*}

\subsection{Systems with constraints on the mass ratio}
The final category considered here consists of three MSPs with
optically bright low-mass companions (Table\,3).  PSRs\,B1957+20 and
J1311$-$3430 \citep{vbk+11,rfs+12,rfc15} belong to a class of
$\gamma-$ray bright eclipsing MSPs with extremely low-mass irradiated
companions.

For PSR\,B1957+20, \cite{vbk+11} derived the mass ratio shown in
Table\,3 after accounting for the fact that due to the strong
irradiation of the companion's surface, radial velocities track the
area facing the pulsar (center of light) rather than the center of
mass.  Using extra constraints on the inclination from the companion's
lightcurve \citep{cvr95}, the pulsar mass at face value is
$2.39(36)$\,M$_{\odot}$. However, \cite{vbk+11} find that the impact
of modeling uncertainties is large and the pulsar mass could be as low
as $1.66$\,M$_{\odot}$.

For PSR\,J1311$-$3430 \citep{rfs+12}, the initial reported value based
on the same technique suggested a pulsar mass with $M >
2.5$\,M$_{\odot}$, but a more recent analysis by \cite{rfc15} shows
that a mass as low as $\sim 1.6$\,M$_{\odot}$ is still possible.

PSR\,J1816+4510 is a binary MSP with an orbital period of $8.7$\,h and
a metal-rich, low mass ($\gtrsim 0.16$\,M$_{\odot}$) companion, the
radial velocity of which implies a high-mass of $m_{\rm p}\sin^3 i =
1.84(11)$\,M$_{\odot}$.

Finally, PSR\,J1740$-$5340 is an eclipsing MSP with a $\sim
0.2$\,M$_{\odot}$ red-straggler companion in the globular cluster
NGC\,6397. This system resembles closely PSR\,J1023+0038 which has
been observed to switch between a rotation- and an accretion-powered
phase.

Given the unresolved discrepancies in the modeling of these systems,
we conservatively assume a randomly oriented orbit and only use the
mass ratio $q$ for our analysis.
\begin{table*}
\begin{center}
\caption{Millisecond pulsar binaries with constraints on the mass ratio}
\begin{tabular}{llrll}
\# & PSR Name & $f(m)$ [M\,$_{\odot}$] & $q$ & Reference \\ 
\hline 
\hline 
1 &J1311$-$3430 &$3\times10^{-7}$  & 175(3) & \cite{rfc15} \\
2 & J1740$-$5340 & 0.002644 & 5.85(13) & \cite{fsg+03} \\
3 & J1816+4510 &    0.0017607& 9.54(0.21) & \cite{kbv+13}           \\
4 &B1957+20   &    $5\times10^{-6}$  & 69.2(8) & \cite{vbk+11}  \\
\hline 
\end{tabular}
\end{center}
\end{table*}
We evaluate the mass of the $j^{\rm th}$ pulsar as: 
\begin{multline}
E_{j}( {\rm data}| m_{\rm p}) = C_j\int dq \int d(\cos i) \times \delta\left[f_{0} - 
f(q,m_{\rm p},i)\right] \\ \times \exp
\left[ - \frac{(q - q_{0}^j)^2}{2\sigma^2_{q_{0}^j}}\right]
\end{multline}
where $q_{0}$ and $\sigma_{q_{0}}$ correspond to the inferred value of
$q$ and its formal uncertainty. The resulting mass likelihoods are
broad and therefore have a small impact on the analysis following
below. In fact, we reach the same main conclusions even if we neglect
these systems entirely.

\section{Statistical Method}
Our main goal is to select the empirical model that best describes the
intrinsic MSP mass distribution.  For each model with a parameter
vector $\pmb{\theta}$, we compute the likelihood as

\begin{equation}
\mathscr{L}({\rm data}|\pmb{\theta}) = \prod_{j}^n \int dm_{\rm p} E_{j}( {\rm data}| m_{\rm p})  
\times P(m_{\rm p}|\pmb{\theta}),
\end{equation}
and then calculate the posterior probability using Bayes' theorem:
\begin{equation}
P(\pmb{\theta} |{\rm data}) = \frac{P(\pmb{\theta}) \times 
\mathscr{L}({\rm data}|\pmb{\theta})}{P({\rm data})}, 
\end{equation}
where  $P(\pmb{\theta})$ is the prior for $\pmb{\theta}$ 
(see next section) and $P({\rm data})$ ensures proper normalization. 

The posterior distribution for the parameter vector $\pmb{\theta}$ is
sampled using a many-particle affine invariant Markov chain Monte
Carlo (MCMC) sampler \citep{emcee2} as implemented in the python
package \texttt{emcee} \citep{emcee}. We experimented with different
number of samplers (from 4 to 800), thinning factors
($0-100$), and initialization strategies. The results were overall
consistent with maximum differences of order $1\%$ in the inferred
marginalized median parameters and the location of the maximum
likelihood in the posterior distribution.  The values reported below
were obtained using 800 samplers, a thinning factor of 50 and 2000
iterations per sampler.  The samplers were initialized in a small sphere
enclosing the preferred model parameters, after some iteration.

 \begin{figure}
\includegraphics[width=0.5\textwidth]{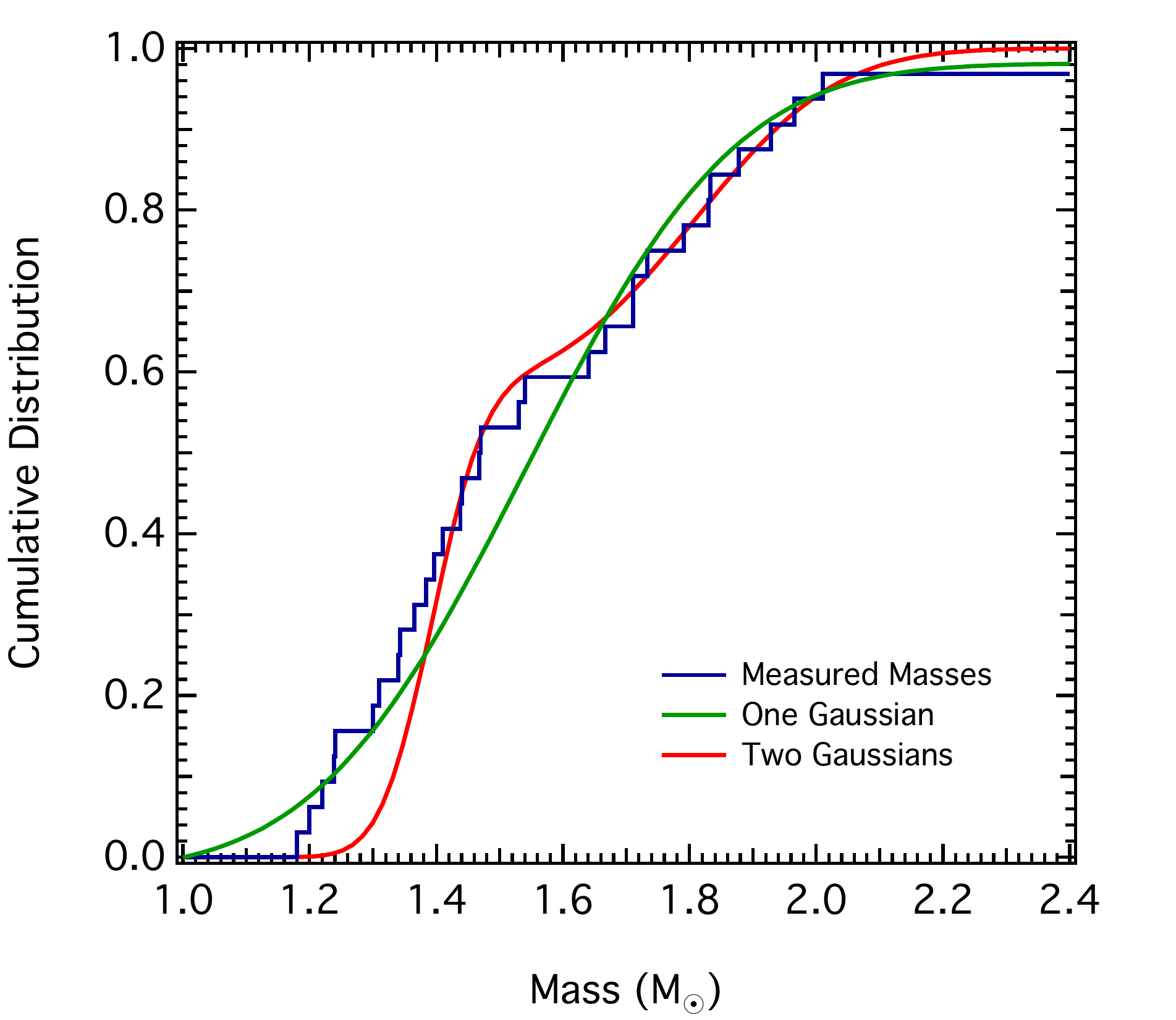}
\caption{Cumulative histogram of MSP masses. The green curve shows the
  cumulative distribution for a single Gaussian, with parameters that
  correspond to the most likely values inferred from the data for this
  intrinsic distribution. The departure of the observed masses from a
  single Gaussian is already evident in this figure and the cumulative
  histogram for the two-Gaussian-component model shown in red provides
  a significantly better description of the data. }
\label{fig:3}
\mbox{}
\end{figure}

\section{Results} 

Before we apply the Bayesian statistical tools to infer the parameters
of the various intrinsic models, we plot the cumulative distribution
of MSP masses to assess visually the level of complexity that we would
need to incorporate into the underlying distributions that can be
supported by the data. In Fig.\,3, we show a
cumulative histogram of the most likely values for the MSP masses. If
the data were described by a single Gaussian, the cumulative histogram
would look like the curve shown in green. However, the presence of
multiple inflection points  strongly suggests the
presence of multiple components in the underlying distribution. It is
evident already from this figure that the two Gaussian component model
shown in red offers a better description of the data. We will now
demonstrate this result quantitatively using the Bayesian inference
method discussed in the previous section.

\subsection{Normal Distribution}
We start by testing the normal distribution (henceforth Model\,I),
\begin{equation}
P(m_{\rm p}| \mu,\sigma) = \frac{1}{\sqrt{2\pi\sigma^2}}\exp\left[ 
- \frac{(m_{\rm p} - \mu)^2} {2\sigma^2}\right]
\end{equation}
as the simplest possible approximation. We implement very weak 
theoretical constraints and use a flat prior with $1.0 \leq \mu \leq
2.5$\,M$_{\odot}$ and $0.0 < \sigma \leq 1.0$\,M$_{\odot}$. 
We also restrict our estimates to $0.8 \leq m_{\rm p} \leq
3.0$\,M$_{\odot}$.

The corresponding posterior probability distribution is shown in
Figs.\,4 and 5.  The posterior samples yield $\mu =
1.542^{+0.054}_{-0.057}$ \,M$_{\odot}$ and $\sigma=
0.260^{+0.061}_{-0.043}$\,M$_{\odot}$ for the median and 68\%
confidence levels (C.L.), which are slightly larger than the values
reported in \cite{opn+12}. This is not surprising as a larger
dispersion is required to accommodate for the  additional high mass
NSs measured after 2012. Nevertheless, this model also predicts  a
high number of stars with $m_{\rm p}\simeq 1.55$\,M$_{\odot}$ and
$m_{\rm p} \le 1.1$\,M$_{\odot}$, which are not observed
(Fig.\,2). Qualitatively, this is a strong indication for asymmetry,
either due to skewness or the presence of a second component at higher
masses.

 \begin{figure}
\begin{center}
\includegraphics[width=0.45\textwidth]{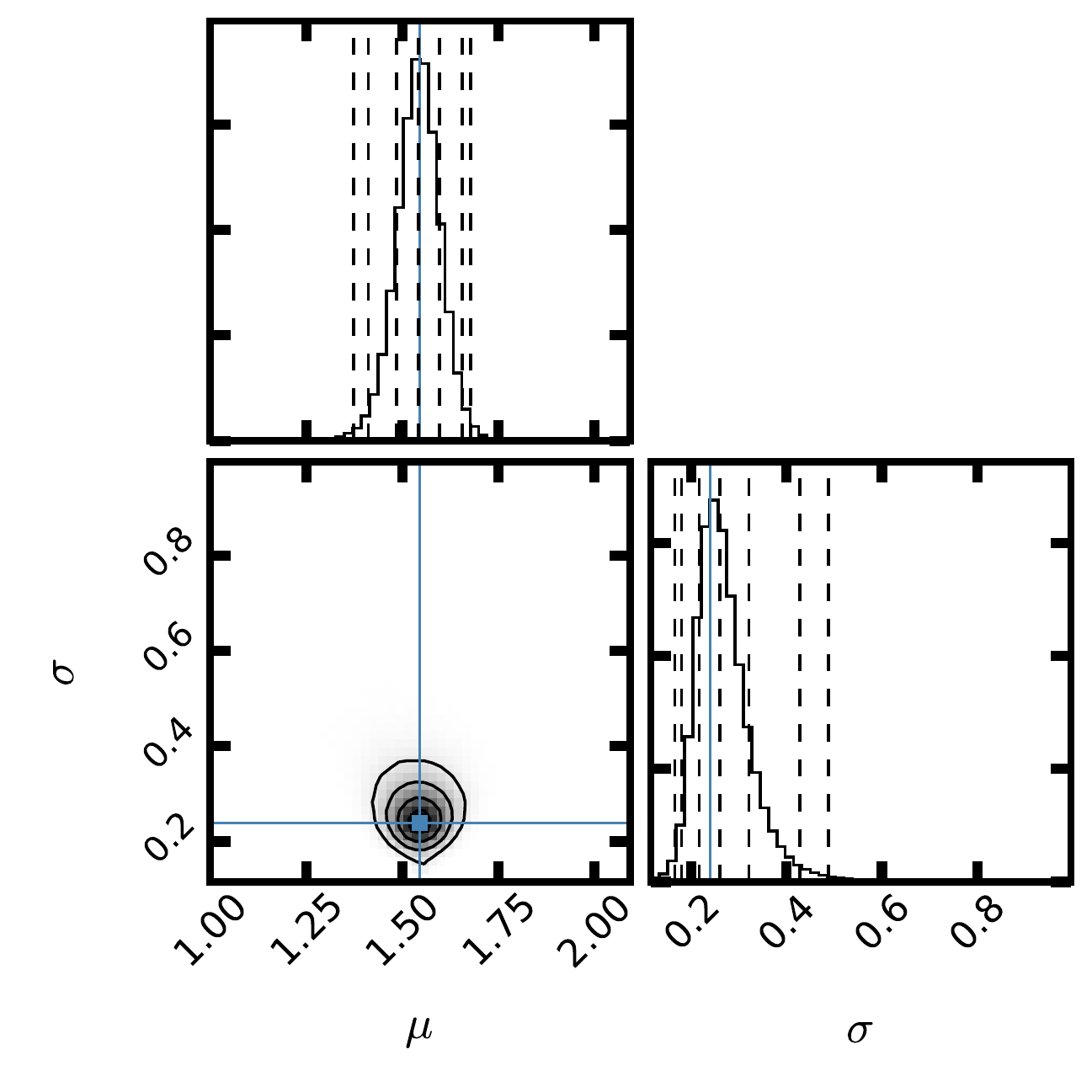}
\caption{Histogram of the MCMC samples drawn from the posterior of the
  normal distribution.  The contours enclose 68, 98 and 99\% of the
  samples. The blue lines show the location of  the maximum likelihood. }
\label{fig:4}
\end{center}
\end{figure}

 \begin{figure}
\begin{center}
\includegraphics[width=0.5\textwidth]{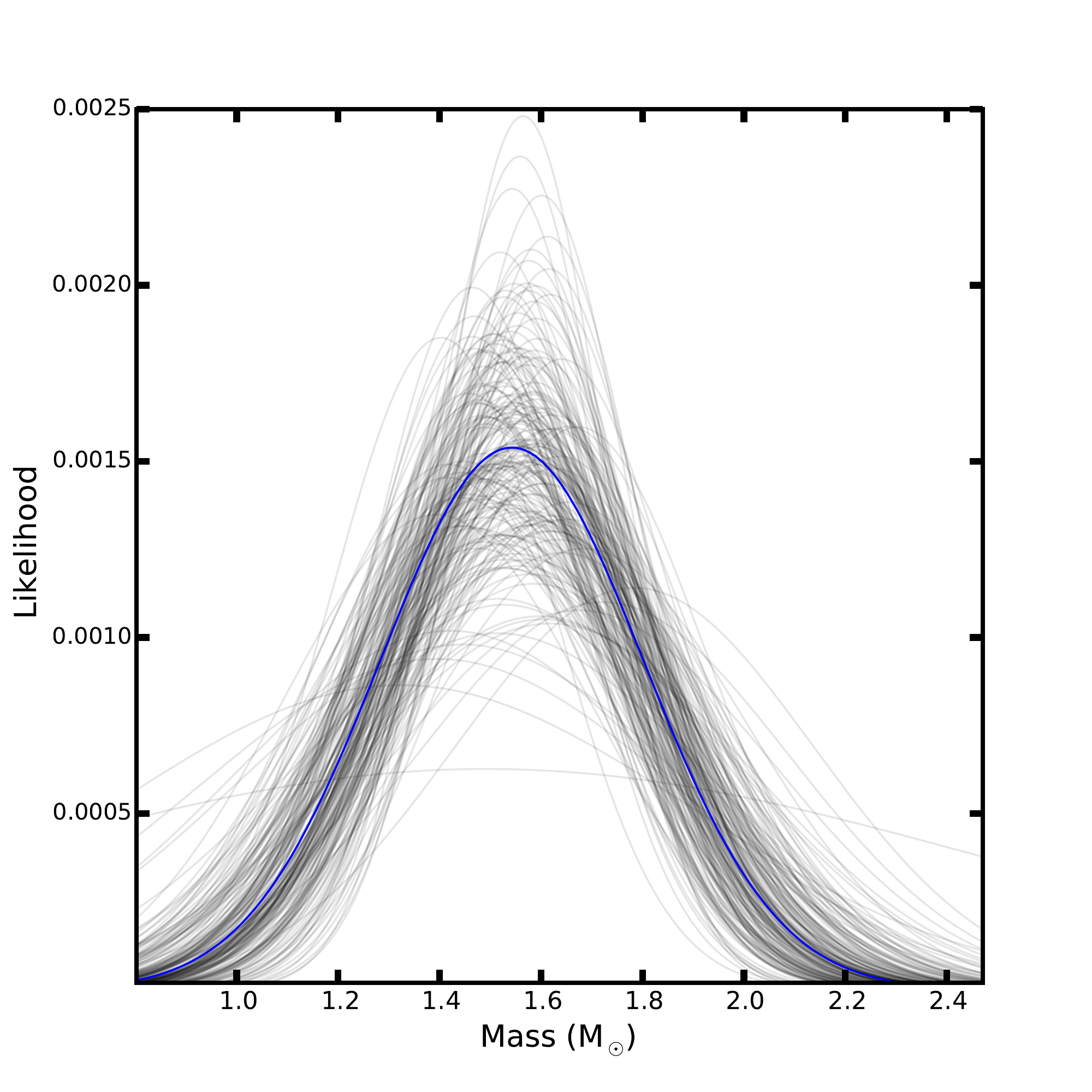}
\caption{Realization of the inferred normal distribution. The blue
  line shows the model corresponding to the median values of the MCMC
  samples. Grey lines represent 1000 samples drawn randomly from the
  posterior and can be viewed as a measure of the uncertainty in the
  inferred parameters. }
\label{fig:5}
\end{center}
\end{figure}

\subsection{Bimodal Normal Distribution} 
We next test for the existence of a second peak using the bimodal
normal distribution (Model\,II),

\begin{multline}
P(m_{\rm p}| \mu_1,\sigma_1,\mu_2,\sigma_2,r) = (1-r)G(\mu_{1},\sigma_{1}) 
+ rG(\mu_2,\sigma_2)
\end{multline}
where $G(\mu_{1,2},\sigma_{1,2})$ are the two normal components and
$r$ is their relative ratio.  One useful property of this model is
that it can also account for skewness in the case of a single- peaked
distribution, e.g., when $\mu_2 = (1 + \epsilon)\mu_1$ and $\sigma_2 >
\sigma_1$, with $\epsilon \in \mathbb{R}$ being a small number.

Here we also use boxcar priors which are set as follows: since the two
components are mutually interchangeable, we use $1.0 \le \mu_{1} \le
1.6$ and $1.45\le \mu_{2} \le 2.8$\,M$_{\odot}$, as well as $0.0<\sigma_{1,2} \le
0.5$\,M$_{\odot}$.  This ensures faster numerical convergence but
still allows for a single-peaked distribution centered around $\sim
1.55$\,M$_{\odot}$, as above.  We adopt $0 \le r \le 1$ for the
relative ratio of the two components.

 \begin{figure*}
\begin{center}
\includegraphics[width=\textwidth]{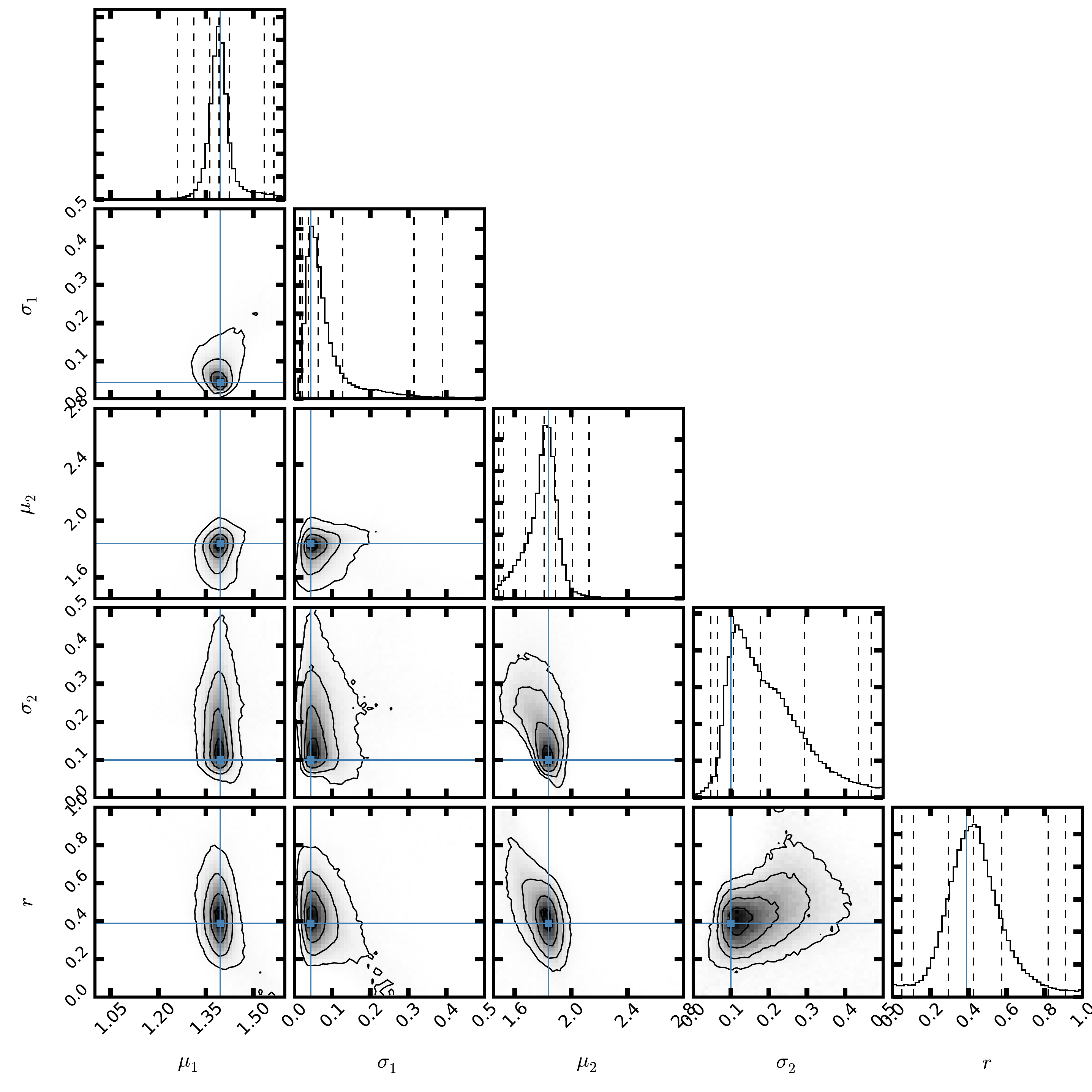}
\caption{Same as Fig.\,2 for the bimodal distribution.}
\label{fig:6}
\end{center}
\end{figure*}

 \begin{figure}
\begin{center}
\includegraphics[width=0.5\textwidth]{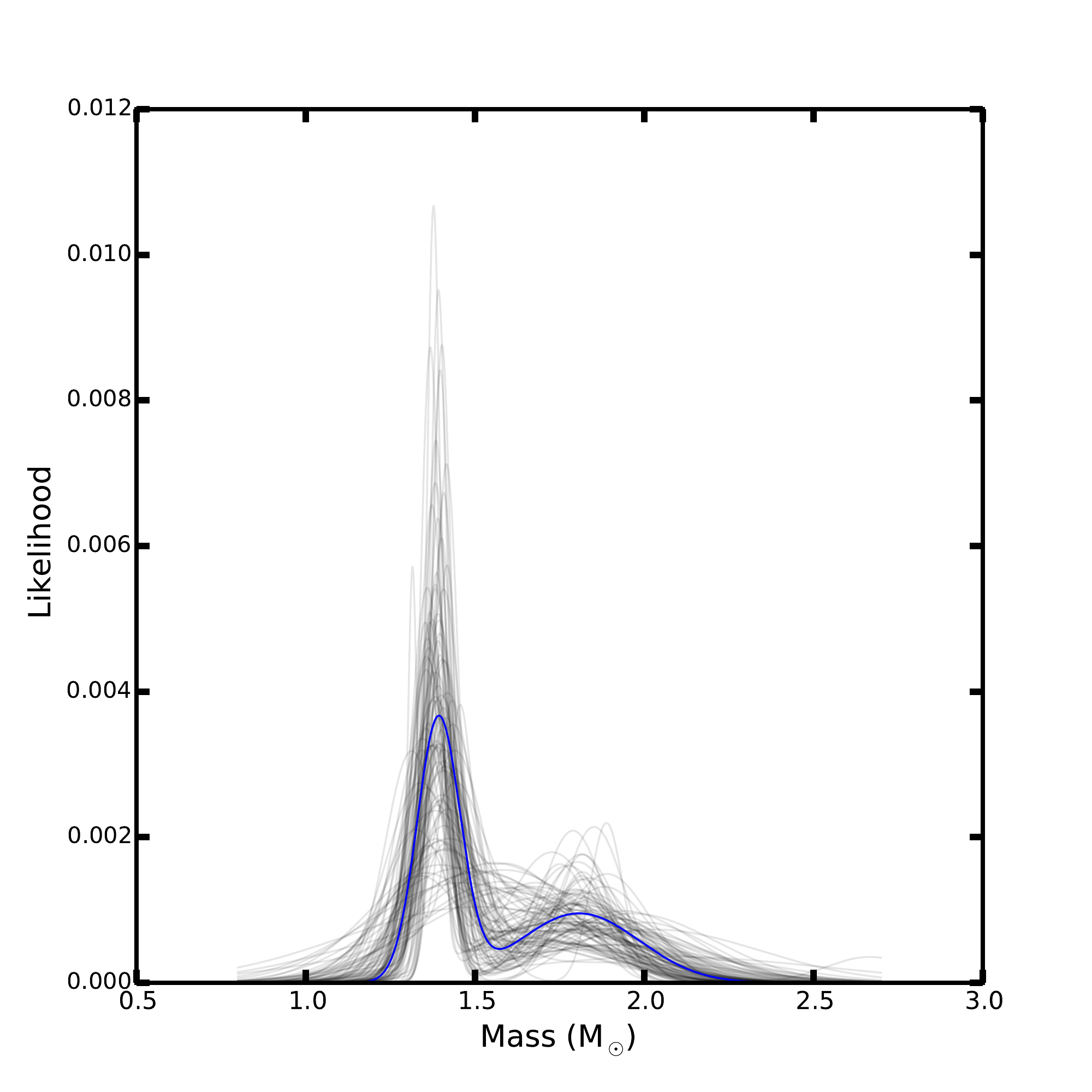}
\caption{Same as Fig.\,3 for the bimodal distribution.}
\label{fig:7}
\end{center}
\end{figure}

The MCMC (Figs. 6 \& 7) yields a maximum likelihood at
$[\mu_1,\sigma_1,\mu_2, \sigma_2,r] =
[1.396,0.045,1.84,0.100,0.389]$.  Overall, for the second peak the
posterior likelihood yields $\mu_{2} = 1.807^{+0.081}_{-0.132}$ and
$\sigma_{2} = 0.177^{+0.115}_{-0.072}$ for the median and 68\% C. L.
Finally, the marginalized posterior likelihood for the relative ratio
of the two peaks has $r =0.425^{+0.150}_{-0.132} $.  A comparison of
the two models follows in the remainder of this section.

\begin{table*}
\caption{Summary of Model Estimates}
 \resizebox{\textwidth}{!}{
\begin{tabular}{lcccccccccccccccc}
     \hline
     & $\mu_1$   &  & & $\sigma_1$ & & & $\mu_2$ & & & $\sigma_2$ & & & $r$ & &   & AICc\\

CL &    $50\%$ & $68\%$ & $98\%$ & $50\%$ & $68\%$ & $98\%$ & $50\%$ & 
$68\%$ & $98\%$ 
& $50\%$ & $68\%$ & $98\%$ & $50\%$ & $68\%$ & $98\%$ & \\
\hline \hline
Model I& 1.542 & 1.485 --1.597 & 1.412 --1.655 & 0.260 & 0.216 -- 0.320 &  0.180 -- 0.427 & -&- &- &- &- 
&- &- &- & - &  461.1 \\
Model II& 1.393 & 1.363 --1.424 & 1.312 --1.535 & 0.064 & 0.038 -- 0.128 &  0.021 -- 0.128 &  1.807 &
1.674 --1.889 &1.518--2.011 & 0.178 & 0.106 -- 0.293 & 0.065 -- 0.436 & 0.425  & 0.293 -- 0.841 & 0.110 -- 0.819 
& 452.4 \\
\hline
     & $\mu$   &  & & $\sigma$ & & & $\log\Delta M$ & & & $\sigma_{\Delta M}$ & & & & &   &\\

\hline
Model III& 1.229 & 1.098 --1.382 & 1.020--1.496 & 0.121 & 0.058 -- 0.228 &  0.042 -- 0.325 &  $-0.514$ &
$-0.885$ -- $-0.272$ &$-1.145$ -- $-0.074$ & 0.332 & 0.172 -- 0.447 &0.036 -- 0.490 & - & - & -
& - \\
\tableline

\end{tabular}
}
\end{table*}

\subsection{Model Selection}
The summary statistics of the two models described above are given in
Table\,4.  Qualitatively, Model\,II seems to provide a better
description of the data as it accounts both for the apparent small
number of pulsars with masses around $\sim 1.55$\,M$_{\odot}$ and
those with $m_{\rm p}<1.1$\,M$_{\odot}$ (see also Figs.\,1 \& 2).

As both models are only  empirical approximations rather
than true physical descriptions of the intrinsic mass distribution, we
employ the second-order Akaike Information Criterion
\citep{akaike,ba02},
\begin{equation}
{\rm AICc}_{i} = - 2 \ln\hat{\mathscr{L}}_{i} + 2k \frac{n}{n-k-1}
\end{equation}
as a means to quantify the relative information loss and select the
model that best describes the data.

The second term in Eq.\,15 introduces a penalty for the complexity of
each model, which depends on the number of free parameters $k$ and the
number of data points $n$.  The relative likelihood can then be
computed as
\begin{equation}
\delta \hat{\mathscr{L}}_{\rm AICc} = \exp\left[\frac{\rm{AIC}_{\rm min} 
- \rm{AIC}_{\rm max}}{2}\right]
\end{equation}
In our case, ${\rm AIC}_{\rm min} =452.4$ for Model\,II and ${\rm
  AIC}_{\rm max} =461.1$ for Model\,I. This yields $\delta
\hat{\mathscr{L}}_{\rm AICc}=0.013$ for Model\,I, which means that
Model II is highly favored, even after accounting for the larger
number of parameters. In the following, we discuss the properties of
Model\,II in more detail.

\subsection{Properties of the Bimodal Distribution}
As briefly mentioned in Section\,4.2, Model\,II does not necessarily
imply the presence of a second component, as it can also account for
skewness in the case of a single-peaked distribution. Indeed, as can
be seen in Fig.~7 and Table~4, the marginalized likelihoods for
$\mu_2$, $\sigma_2$ and $r$ span a wide range of values, even allowing
for a normal distribution peaking around $1.55$\,M$_{\odot}$ within
the 99\% C.L. To quantify the separation of the two components, we use
the statistic \citep{ashman},
\begin{equation}
D = 2^{1/2}\frac{|\mu_1 - \mu_2|}{\sqrt{\sigma^2_1 + \sigma^2_2}}
\end{equation}
According to \cite{ashman}, for a mixture of two normal distributions,
a clean separation requires $D>2 $.  The histogram for $D$ inferred
from the posterior is shown in Fig.\,8. We find $D =
3.12^{+1.95}_{-1.61}$, with $D \ge 2$ for $ 73\%$ of the samples.
Hence, albeit less favored, a uni-modal distribution with a
significant positive skewness cannot be ruled out conclusively with
the existing sample of MSP masses.

 \begin{figure}[h]
\begin{center}
\includegraphics[width=0.5\textwidth]{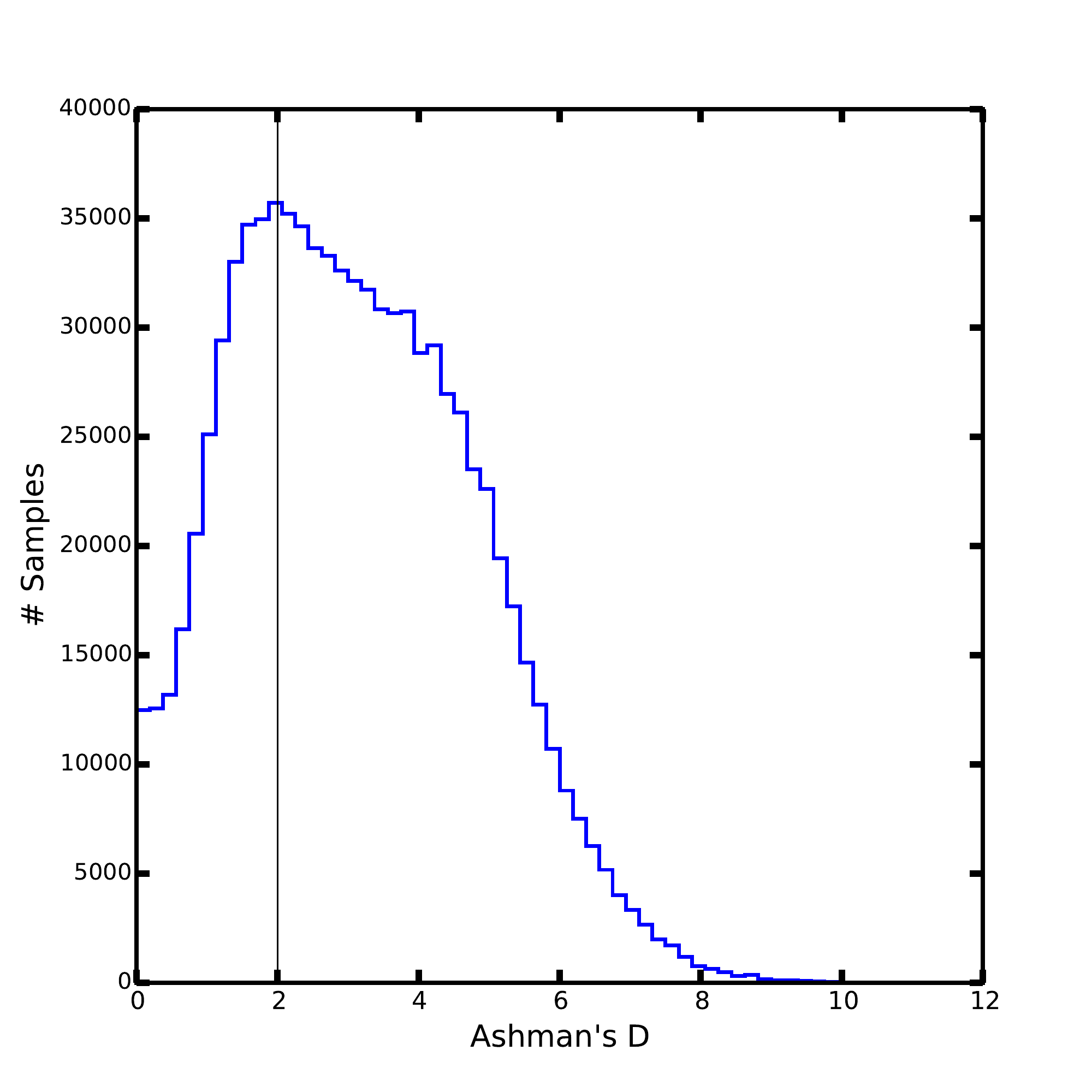}
\caption{Histogram of the Ashman-D statistic for the MCMC samples
  drawn from the bimodal distribution. 73\% of the samples have $D\geq2$, favoring a bimodal distribution}
\label{fig:8}
\end{center}
\end{figure}

An alternative way to assert the likelihood of an asymmetric uni-modal
distribution is to test whether mass accretion onto the NS can
reproduce the observed sample, starting with normally distributed
birth masses. To do this we assume that the masses of recycled pulsars
can be described by,
\begin{multline}
P(m_{\rm p}|\mu,\sigma_{\mu}) = \\ \int d{\Delta M}\int 
d\sigma_{\Delta M}P_{\rm b}(\mu - \Delta M,\sigma_{ M}) P_{\rm a}(\Delta M,\sigma_{\Delta M}), 
\end{multline}
 which we refer to as Model\,III in what follows. Here $P_{\rm
  b}(\mu-\Delta M)$ follows Eq.\,13 and represents the MSP birth-mass
distribution, and
\begin{multline}
P_{\rm a}(\Delta M) = \frac{1}{\Delta M \ln(10) 2\pi (\sigma^2_{\Delta
    M})^{1/2}} \times \\ \times \exp\left[ - \frac{(\log \Delta M -
    \log \Delta M_{0})^2}{2\sigma^2_{\Delta M}}\right]
\end{multline}
is a log-normal distribution which we use to approximate the effect
of mass accretion.  We show the postrior likelihoods for the
parameters of Model\,III in Fig.\,9.  Unlike Models\,I \& II, the
posterior is not well-localized and consequently, the selection
criteria used above cannot be applied directly. However, a qualitative
comparison with Models\,I and II is still possible by examining the
physical implications of the maximum likelihood model, located at
$[\mu-\Delta M,\sigma_M, \log_{10} \Delta M, \sigma_{\Delta M}] = [1.23,0.040,-0.511,0.464]$. 
The ``birth mass'' distribution $P_{\rm b}$ -- likely constrained by 
low-mass NSs close to the range $1.2-1.4$\,M$_{\odot}$ -- is narrowly distributed 
around an extremely low mass. This would imply that all massive NSs were born 
with an extremely low mass and subsequently accreted a substantial amount of material,  
$\Delta M$. As we discuss below in more detail, this assertion is 
disfavored by stellar evolution considerations in at least 
a subset of the cases, supported by the systems' observed
properties (for instance, the nature of their companions). A more
plausible model would require a relatively broad birth-mass
distribution, and an accretion kernel that yields high probabilities
for small $\Delta M$s. However, as can be seen in Figure\,8, such
models have low likelihood values compared to the preferred model.
For this reason, in any further calculations, we focus only on 
Models\,I and II.

\begin{figure}[h]
\begin{center}
\includegraphics[width=0.4\textwidth]{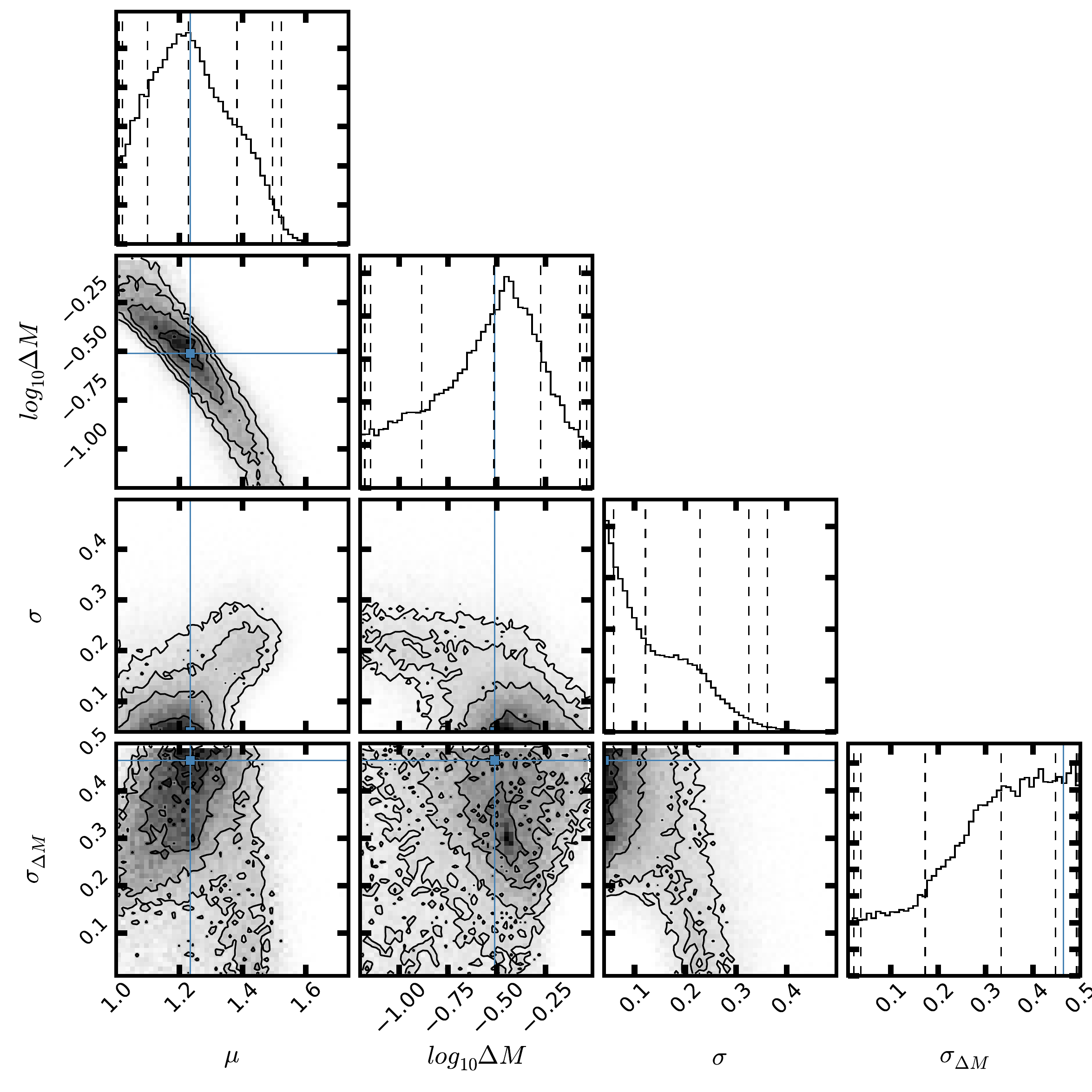}
\caption{Posterior likelihoods over the parameters of a model that
  convolves a Gaussian birth-mass distribution with a log-normal
  kernel that represents the effects of accretion (Model\,III; see
  Section\,4.4).}
\label{fig:9}
\end{center}
\end{figure}

\begin{figure*}[h]
\begin{center}
\includegraphics[width=\textwidth]{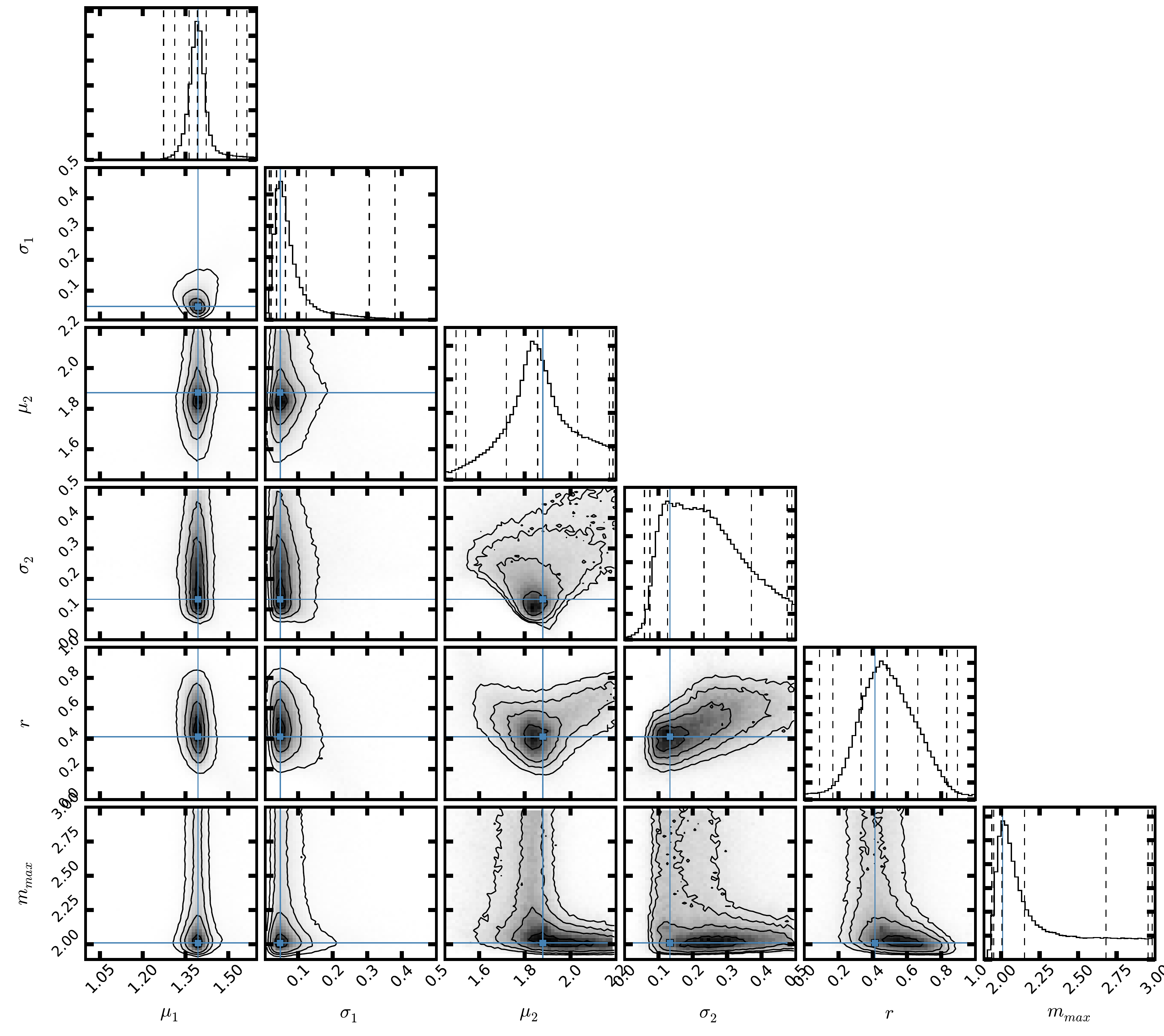}
\caption{Same as in Fig.\,4 for the truncated bimodal distribution.}
\label{fig:10}
\end{center}
\end{figure*}

\section{Constraints on the Neutron Star Equation of State}
One interesting implication of the inferred distribution is that
$~3\%$ of NS must have masses above $2.1$\,M$_{\odot}$.  However, the
true predictive power of the model is small, as the existence of such
high-mass NSs depends primarily on the underlying EoS.  Indeed, there
exist compelling theoretical evidence in favor of a relatively soft
EoS, which can only support NSs with masses below $\le
2.1$\,M$_{\odot}$. This would introduce a high-mass cut off in the
distribution.  With the framework developed here, we can place
constraints on the maximum NS mass by including an extra truncation
parameter, $M_{\rm max}$ so that
\begin{equation}
P(m_{\rm p}| \pmb{\theta}) =\begin{cases}
P(m_{\rm p}| \mu_1,\sigma_1,\mu_2,\sigma_2,r), & \text{if $m_{\rm p}\le M_{\rm max}$}\\
0, & \text{otherwise}.
\end{cases}
\end{equation}
The posterior probability of this model is shown in Fig.\,10 and  yields  
\begin{equation}
M_{\rm max} \ge 2.018\,{\rm M}_{\odot}\,\text{(98\% C.L.)},
\end{equation} 
or $M_{\rm max} \ge 1.924\,{\rm M}_{\odot}$ at 99.98\% C.L..
This is imposed mainly by the massive pulsar PSR J0348+0432.  Interestingly,
this limit appears to be almost insensitive to the choice of the
underlying model for $m_{p} \le M_{\rm max}$; switching from Model\,II
to I yields $M_{\rm max} \ge 2.019\;{\rm M} _{\odot}$ (98\% C.L.).

\subsection{Expectations for Future Pulsar Surveys}
The detection of a high-mass truncation would be of high 
importance for nuclear physics calculations as it would provide direct
constraints on the EoS at very high densities \citep[see]{op09}.  The
inferred MSP mass distribution and Eq.\,20 allow  to assess the
likelihood of measuring $M_{\rm max}$ in the future.

Pulsar surveys planned for, e.g., the Square-Kilometre Array (SKA) in
Phase\,2, will provide a nearly complete census of the radio-pulsar
population in the Galaxy \citep{kbk+15}. Follow-up observations of
these discoveries are expected to yield precision mass measurements
for over 350\,MSPs \citep[e.g.][ and references
  therein]{tkb+15,agp+15}.  Using our estimates above, we can simulate
their underlying mass distribution, include a cut-off at a given
$M_{\rm max}$, and then use Eq.\,20 to assess its detectability.

The robustness of such a simulation depends critically on a realistic
estimate for the expected precision of NS mass-measurements with
future instruments, which is beyond the scope of this work.  For
masses derived with radio timing, it is safe to assume that the
distribution of uncertainties will remain nearly constant. This is
because both the survey sensitivity and timing precision scale
similarly with the telescope size.  For optical spectroscopy,
uncertainties depend on the companion's brightness and therefore for a
given telescope the achieved precision will scale (at least) with the square of
the distance.  Nevertheless, future optical telescopes such as LSST,
E-ELT and TMT will significantly increase the Galaxy volume that can
be probed with this technique.
 
For the estimates presented here we adopt a distribution of
measurement uncertainties that is uniform between the minimum
and maximum errors shown in Table\,1.
\begin{figure}[h]
\begin{center}
\includegraphics[width=0.5\textwidth]{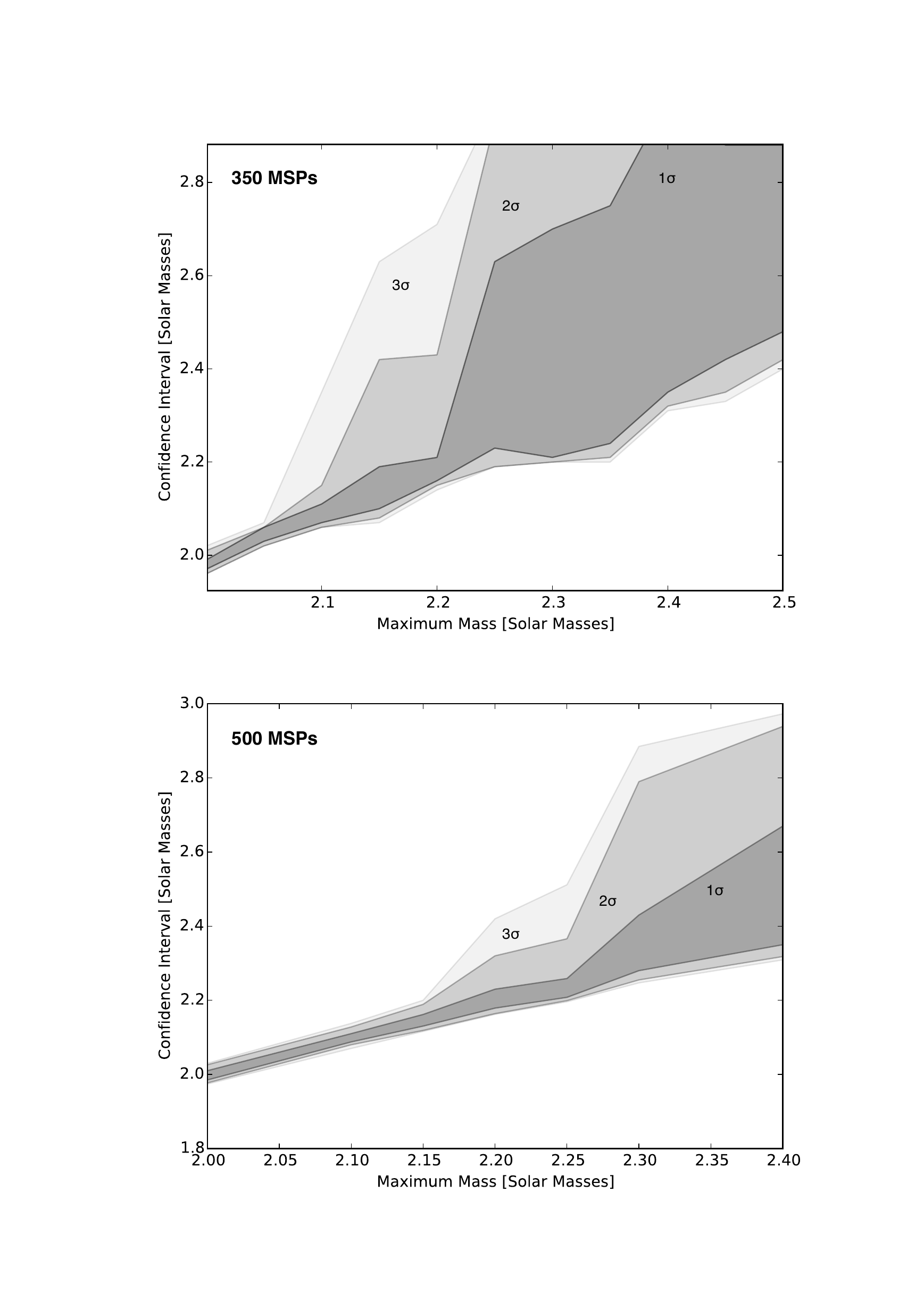}
\caption{Projected precision on the localization of the maximum NS
  mass as a function of the number of MSP mass measurements.}
\label{fig:11}
\end{center}
\end{figure}

For our simulation, we first use the median parameters for Model\,II
introducing a cut-off mass $M_{\rm max}$ at values ranging from $2.0$
to $2.5$\,M$_{\odot}$. We then draw a number of MSPs to which we
assign uncertainties that follow the aforementioned uniform
distribution. Finally, we apply Eq.\,20 and examine the marginalized
posterior probability for $M_{\rm max}$.

We show the results of two simulations for 350 and 500 MSPs  in
Fig.\,11. In the former case, a high-mass truncation would be
detectable up to $\sim 2.15$\,M$_{\odot}$ with a $3\sigma$ precision
of $\sim0.2$\,M$_{\odot}$. In the second scenario, $M_{\rm max}$ can
be detected up to 2.25\,M$_{\odot}$, a value that is above the
predictions of a large number of EoS models.

\subsection{Massive NSs in the known MSP population}
Another  question related to the maximum mass is whether if we can identify massive NSs among MSP
systems that do not have any measured constraints on their total mass
or mass ratio.  There are currently $\sim 200$ binary MSPs listed in
the ATNF pulsar catalogue\footnote{http://
  www.atnf.csiro.au/research/pulsar/psrcat/} \citep{mhth05}, excluding
those listed in Tables 1-3.

Based on our models, $\sim 40$ of these should have masses above
$1.8$\,M$_{\odot}$. However, these systems are impossible to identify
directly without any further information.  One possibility is to
search among those systems that may have experienced a significant
episode of mass transfer, e.g., those associated with eclipsing
$\gamma-$ray systems (see Section\,2). As we briefly discussed in the introduction,
there is indeed accumulating evidence for massive NSs in these
binaries.  Unfortunately, precision measurements are hard to achieve
with current methods, as the timing precision is relatively poor and
the pulsar companions are not understood well enough to provide
meaningful mass estimates using optical spectroscopy. Nevertheless,
alternative methods such as those based on pulsar scintillometry
\citep{pmd+14} may help to overcome these limitations in the future.

Another approach is to employ constraints imposed by binary evolution
theory.  For example, the masses of He-core white dwarf companions to MSPs are
known to correlate tightly with the orbital period, (e.g.,
\citealt{ts99a}; see also Fig.~2 in \citealt{tv14} for a comparison
with recent data).  The main reason for this is the correlation
between the radius and the degenerate core mass of the (sub)giant star
progenitor of the He~white dwarf. Hence, from the measured orbital period
$P_{\rm b}$, we can determine the mass of the He~white dwarf, $M_{\rm WD}$, and
thus, given the mass function of the observed system, obtain the mass
of the pulsar as a function of orbital inclination angle.

\begin{figure}[t]
\begin{center}
\begin{tabular}{c}

\includegraphics[width=0.4\textwidth,angle=270]{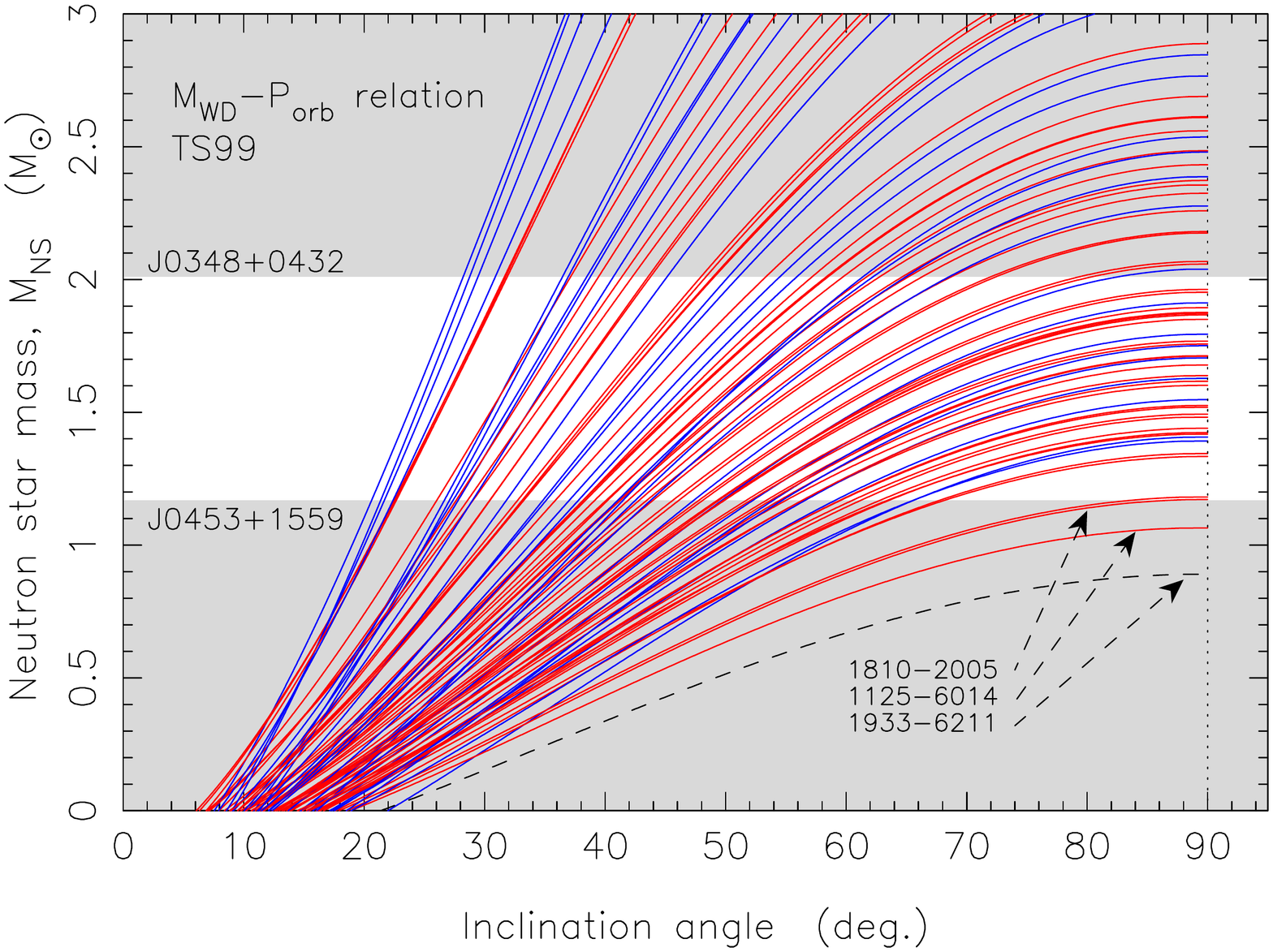}\\
\includegraphics[width=0.4\textwidth,angle=270]{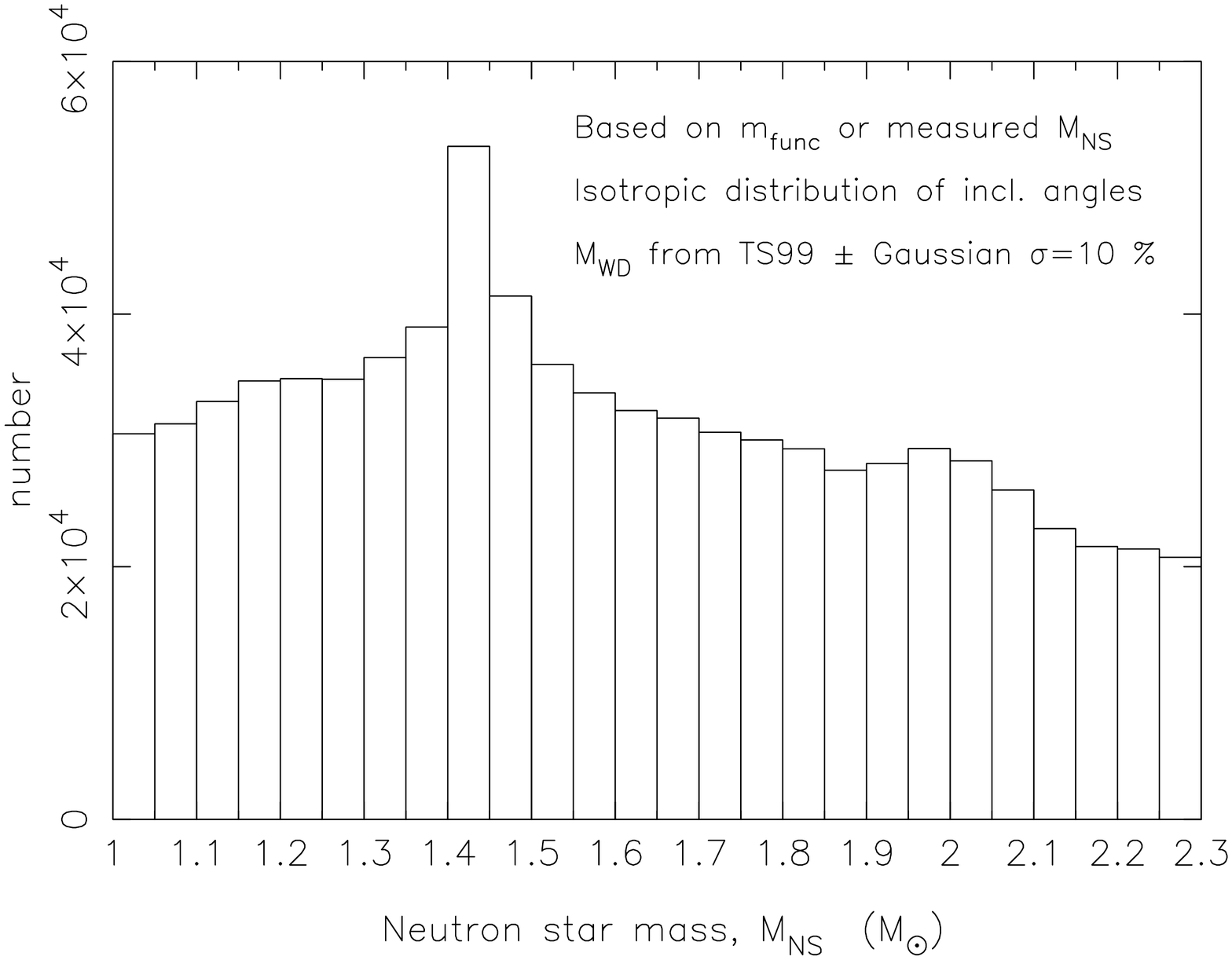}
\end{tabular}
\caption{NS masses as a function of inclination (upper panel) and
  accumulated mass likelihoods (lower panel) for all known MSPs with
  He~white dwarf companions, based on the $M_{\rm WD}-P_{\rm b}$~relation of
  \cite{ts99a}. The summary statistics can be found in Table\,5.  The
  red curves are for binaries with $P_{\rm b}<30\;{\rm days}$; blue
  curves are for systems wider than 30~days.  PSR~J1933$-$6211 (black
  curve) is an example of a system with a CO~white dwarf which does not
  obey the $M_{\rm WD}-P_{\rm b}$~relation, and thus it is not
  included in our statistical sample.}
\end{center}
\end{figure}

The top panel of Fig.\,12 (the '{\it spaghetti plot}') shows this
result for all known MSPs with He~white dwarf companions, calculated using
the relation of \citet{ts99a} for solar chemical composition.  The red
curves are for binaries with orbital periods of less than 30~days;
blue curves are for systems wider than 30~days.  There is a slight
trend for wider systems having marginally more massive NSs (see
discussions below).

Assuming an isotropic distribution of orbital inclinations as in
Sections\,2.3.2 \& 2.3.3, we can calculate the mass likelihood for
each pulsar as:
\begin{multline}
\mathscr{L}_{j}( m_{\rm p}| m_{\rm c}) = C_j\int dm_{\rm c} \int d(\cos i) 
\times \delta\left[f_{0} - f(m_{\rm c},m_{\rm p},i)\right] \\ \times \exp
\left[ - \frac{(m_{c} - m_{0}^j)^2}{2\sigma^2_{m_{0}^j}}\right],
\end{multline}
where $m_{0}$ is the theoretical prediction for the white dwarf mass. We show
in the bottom panel of Fig.\,12 the expected NS mass distribution
marginalized over the inclination angle based on these likelihoods for
all known MSPs with He~white dwarf companions. Here, we calculated the white dwarf mass
using the $M_{\rm WD}-P_{\rm b}$~relation of \cite{ts99a}, assumed a
fractional uncertainty of $10\%$ (see also Table\,5 in the Appendix),
and assumed that there are no selection effects in the observed
sample. Only in cases where the MSPs mass is measured (cf. Table\,1),
we use this value and its associated uncertainty.  We disregarded the
solutions for NS masses outside the interval of
$1.0-2.3$\;M$_{\odot}$.

At first sight, there seems to be a significant number of potentially 
massive NS candidates. Nevertheless, it should be stressed that these
results depend on the applicability of the $M_{\rm WD}-P_{\rm
  b}$~relation. For example, some systems may simply reflect a
violation of required assumptions behind this relation, i.e., the
evolution in LMXB systems with donor star masses $\la
2.3$\;M$_{\odot}$ \citep{ts99a}.  An example of an MSP which did not
evolve from an LMXB system is PSR~J1933$-$6211 (plotted as a dashed black
curve in the top panel of Fig.\,12).  This white dwarf companion has a CO~core 
(Matteo {\it et al.}, in prep.) and, thus, the system most likely
evolved from an intermediate mass {X}-ray binary \citep[like PSR~J1614$-$2230,][]{tlk11}, i.e.,
under conditions in which a non-degenerate core developed in the white dwarf
progenitor star. Therefore, this system will not obey the $M_{\rm
  WD}-P_{\rm b}$ relation, which would have predicted an unrealistic
NS mass of $<0.9$\;M$_{\odot}$. Interestingly enough, in the bottom
panel of Fig.\,12, a peak of a NS mass near $1.4$\;M$_{\odot}$ is seen
in this distribution of MSP masses, similar to that shown in Fig.\,7.

\section{Discussion}
\subsection{Summary}
In this work, we used the available ensemble of MSP mass measurements
to examine the underlying recycled NS mass distribution. The main
results can be broadly summarized as follows:
\begin{itemize}
\item A normal distribution does not seem to provide an adequate
  description of the observed MSP masses. Based on our Bayesian
  analysis method outlined in Section\,3, we find strong evidence for
  asymmetry. More specifically, a bimodal distribution peaking at
  $\sim1.4$ and $\sim1.8$\,M $_{\odot}$ is favored by the data, but a
  single-peaked distribution with strong positive skewness is allowed
  within 20\% of the posterior likelihood and cannot be conclusively
  ruled out (see Section 4.3).

\item Massive NSs seem to be more common than previously thought. In the inferred distributions, we
  find that approximately $20\%$ of binary MSPs have masses above
  $1.8$\,M$_{\odot}$.  This number is independent of the assumption
  made for the shape of the underlying mass distribution.

\item Including a high-mass truncation in our models yields a robust
  estimate on the maximum NS mass of $M_{\rm max} \ge
  2.018$\,M$_{\odot}$ at 98\% C.L. This result is again not sensitive
  to the adopted distribution model.

\item As the number of mass measurements increases, it may become
  possible to precisely measure a maximum mass cut-off.  More
  specifically, our simulation shows that with 350 MSP mass
  measurements following the currently favored bimodal distribution,
  it will be possible to localize an $M_{\rm max}$ lower than
  2.15\,M$_{\odot}$ with a precision better than 5\%.
\end{itemize}

\subsection{Selection Effects}
Before discussing the ramifications of these results in more detail,
it is necessary to consider possible selection effects that may bias
our findings.

First, our dataset is intrinsically biased in the sense that it only
contains NSs in binary systems.  As binarity can significantly affect
the outcome of massive star evolution \citep{lan12}, the inferred mass
distribution may differ substantially from that of isolated NSs.
Currently, there exist very little information for the masses of
single NSs.  A recent study of micro-lensing events toward the
Galactic Bulge in the OGLE\,III survey \citep{wks+15} suggests that
the data are consistent with a uniform mass distribution of single
compact objects but the uncertainties in the individual mass
measurements using this method are still very large. Furthermore, it
is possible that some of these lensing sources are normal stars
located close to the lensed background objects.  The ongoing Galactic
survey by GAIA \citep{be02}, as well as the 4th phase of the OGLE
survey \citep{uss15} and the KMTNet experiment \citep{kmtnet} will significantly improve on the aforementioned
result and may even make it possible, for the first time, to identify
photometric or astrometric micro-lensing events caused by radio
pulsars \citep{dsl+15}.  If the distance and transverse velocity of a
lensing pulsar can be determined, then it may become possible to
measure its mass with a precision of order 10\% \citep{dsl+15}.

A second bias when discussing NS masses is that in this work we have
only considered MSPs, most of which have likely experienced a
long-term period of mass accretion from their binary companion. The
extent to which accretion shifts the current masses away from the NS
birth masses depends purely on the ability of the NS to accumulate
mass during the LMXB phase. As we will argue in more detail below,
observational evidence suggest that, in most cases, the accretion efficiency is
 low, possibly smaller than $\sim 10\%$.  Hence, even though
it is not possible to completely deconvolve the impact of mass gain,
MSP masses can still serve as a meaningful probe of the SN mechanism
and the structure of their progenitors, as we discuss in more detail
in Section\,6.4.

Another point of caution is the small number of MSPs employed in the
analysis: we use 32 MSPs of which only 18 have precisely measured
masses.  This makes our result sensitive to small number statistical
fluctuations. For instance, the identification of the high-mass peak
component in the distribution of MSP masses is only possible due to
the recent precision measurements of a few high-mass NSs like PSRs
J0348+0432 and J1949+3106, in addition to PSR\,J1614$-$2230, already
considered in previous studies by \cite{opn+12} and
\cite{kkd+13}. Even though these measurements are robust, a
measurement bias not reflected in the formal uncertainties may alter
 the posterior probability distribution, in particular
the location and dispersion of the high-mass peak component.

Finally, one must also consider possible selection effects in the MSP
sample itself, in particular those stars with precise mass
measurements (Table\,1).  Masses derived from Shapiro delay are easier to achieve in short orbital period binaries with high
inclinations. For longer period binaries, the signal is weaker but it
also becomes logistically more difficult to achieve the required observing
cadence. Similarly, the optical spectroscopy method is more relevant
to compact systems. Here, fine orbital sampling and homogeneity are
equally important, but there also exists an anti-correlation between
the binary orbital period and the intrinsic luminosity of the white dwarf
companion. This is because the stellar envelope thickness, which
determines the cooling rate, depends sensitively on the white dwarf mass, which
in turn scales with the orbital separation.  Nevertheless, within our
sample, there is no evident correlation between the mass of a binary
MSP and its direct observational properties.  For instance, the
relativistic pulsars PSRs J1738+0333, J0751+1807 and J0348+0432 have
similar spin and orbital properties, but very different masses.  An
exception to this is possible for the eclipsing MSPs in Table\,3 which
may be massive as a class, although any conclusion is premature with current data. Here however, we only used their mass
ratios in our analysis, which results in very broad likelihoods for
their masses. Omitting this class entirely leads to almost
identical posterior likelihoods and, therefore, does not impact any of
the conclusions.

\subsection{Asymmetry in the MSP mass distribution: Accretion vs NSs born massive}
One of our key results is the detection of asymmetry in the MSP mass
distribution, most likely caused by the presence of a high mass
component centred around $1.8$\,M$_{\odot}$.

As briefly discussed above, one mechanism to produce an asymmetric
mass distribution is accretion from a binary companion. In
Section\,4.4, we tested this hypothesis using a simple empirical model
to simulate the effect of mass accretion. We found that the highest
likelihood model implies that all NSs, including those with high
masses, had birth masses close to $1.1-1.3$\,M$_{\odot}$.  If this
were the case and the birth masses of MSPs were similar, then we would
expect a tight correlation between the observed masses and companion
type, which is a direct probe of the magnitude and duration of the
mass exchange episode. This is clearly not the case.  For instance,
PSR\,J1614$-$2230 has a CO~white dwarf companion and most likely formed via a
Case A Roche-lobe overflow from at least a $\sim 4.0$\,M$_{\odot}$ main
sequence donor. Consequently, the NS in this system did not experience
a long-term accretion episode and, therefore, its birth mass must have
been at least $1.7$\,M$_{\odot}$ \citep{tlk11,lrp+11}.  Similarly,
systems like PSRs\,J1918$-$0642 and J1738+0333 have very low NS masses,
but otherwise appear to be fully recycled.

Hence, we conclude that the high number of massive NSs most likely
reflects differences in the NS birth masses and that at least some of
them must have been born with a mass larger than
1.6\,M$_{\odot}$. This again connects back to the core-collapse
mechanism and the properties of the high-mass progenitor stars prior
to the SN explosion.

Possibly the largest determining factor for the mass and nature of the
compact remnant is the size of the progenitor's iron core at the onset
of core collapse.  The iron core mass depends on whether carbon
burning proceeds convectively or radiatively, and thus it is sensitive
to the $^{12}{\rm C}/^{16}{\rm O}$ ratio at the depletion of central
helium burning.  This ratio depends primarily on the initial stellar
ZAMS mass and whether the star evolves in isolation/wide-orbit binary
or in a close binary system (\citealt{wl99,bhl+01,plp+04}; see
\citealt{tlk11} for a summary).  

In close binaries, which are more
relevant to our MSP sample, the progenitor star of the NS may lose its
hydrogen-rich envelope at an early stage, thereby reducing the growth
of its helium core, resulting in a larger $^{12}{\rm C}/^{16}{\rm O}$
ratio -- similar to what is expected for the lower-mass end of massive
isolated stars.  In more massive stars ($M>20$\,M$_{\odot}$, or
somewhat less massive stars evolving in isolation), however, the
destruction of carbon via the $^{12}{\rm C}(\alpha,\gamma)^{16}{\rm
  O}$ reaction tends to dominate over its creation via the $3\alpha$
process.  Therefore, the net central carbon abundance is lower,
leading to relatively fast radiative carbon burning, leaving behind a
high central entropy and production of a more massive iron core.  A
large fraction of these stars are expected to form BHs via
fallback. However, it is possible some of them will produce high-mass
NSs instead, depending on the degree of stripping of envelope mass
\citep{tlp15}, the SN explosion physics \citep{ujma12,pt15,mhlc16} and the
EoS.  Other factors such as metallicity, stellar winds, rotational
mixing and angular momentum transport, B-fields, and the location of
the outermost oxygen burning shell, may influence the nature of the
final remnant as well \citep{whw02,hfw+03}.

\subsubsection{Further evidence for NSs born massive}
In the previous section we interpreted the diverse characteristics of
high-mass MSPs as a strong indicator for NSs born massive. Further
evidence supporting this claim can be found in other binary NS types,
where the effect of mass accretion is less severe.  For instance, some
high-mass {X}-ray binaries (HMXBs) like Vela~X$-1$ \citep[$M_{\rm NS}
  = 2.1(1)$\,M $_{\odot}$;][]{fbl+15} and 4U\,1700-37 \citep[$M_{\rm
    NS} = 2.4(3)$\,M$_{\odot}$;][]{kvvv06} may host massive NSs. For
these systems, the maximum mass that could have been accreted from
their companion is bound by the combination of a short evolution
timescale of their high-mass donor ($\tau_{\rm nuc}\sim 10$\,Myr) and
the Eddington limit for accretion onto a NS ($\dot{M}_{\rm Edd} \simeq
{\rm ~a~few}\;10^{-8}$\,M$_{\odot}$\,yr$^{-1}$).  In reality, the
actual accretion rate is likely much smaller, as evident by the
relatively low X-ray luminosities of most known HMXBs \citep{ggs02},
which imply that the total amount of accreted mass probably does not
exceed $\Delta M\simeq 0.01$\,M$_{\odot}$ (Tauris~et~al., in prep.).

DNSs are another example of binaries where mass accretion onto the NS
is not important for their final mass.  All known DNSs host NSs with
masses between $1.23-1.44$\,M$_{\odot}$ with the exception of the
recently discovered PSR\,J0453+1559 \citep{msf+15}, in which a
1.56\,M$_{\odot}$ NS orbits a $1.17$\,M$_{\odot}$ companion.  While
there can be several episodes of mass transfer between the HMXB and
DNS phases, the net mass gain on the first born NS is likely to be
less than $\sim\!0.01$\,M$_{\odot}$, part of which is accreted during
the common envelope phase.  Consequently, if PSR\,J0453+1559 is indeed
a DNS, this $1.56\;M_{\odot}$ pulsar must have been born with a mass
close to the observed one.
 
\subsection{Long-term accretion efficiency in LMXBs}
Assuming that the intrinsic MSP mass distribution is indeed bimodal,
another insightful finding is the small dispersion  for the low-mass
component, implied by the posterior likelihood. This is similar to
that inferred for DNSs by \cite{opn+12}. All of the low-mass MSPs in
our sample appear to be fully recycled, as evident by their companion
types, spin periods and magnetic fields.  The ZAMS progenitors of the
white dwarf companions had initial masses between $\sim1.0$ and
2.3\,M$_{\odot}$ \citep[see][and references
  therein]{avk+12}. Therefore, the total mass transfer during the LMXB
phase was of order $0.6-2.1$\,M$_{\odot}$ to generate the observed
$\sim 0.16-0.4$\,M$_{\odot}$ white dwarf companions. Efficient accretion would
imply that most of the NSs in these systems had initial masses below
$1.0$\,M$_{\odot}$, which is highly unlikely, and such mass transfer
(if close to conservative) would also produce a larger dispersion in
the NS masses than seen in the observed data.  Therefore, it is again
likely that the birth masses of the MSPs were not much different from
those observed today, typically smaller by $\le 0.1\;M_{\odot}$. We
note that such a relatively small amount of accreted material is
indeed sufficient to recycle the MSPs to spin periods of a few ms
\citep{tlk12}.  Our conclusion \citep[see also][]{avk+12} that the
accretion during the LMXB phase is highly inefficient, corresponding
to average accretion efficiencies of only $\sim 5-20\;{\rm \%}$, means
that the far majority of the transferred matter, despite sub-Eddington
mass-transfer rates in many cases, is lost from the LMXB system via
accretion disk instabilities \citep{vpa96,dhl01} and/or propeller
effects (see, e.g., \citealt{is75}).

\subsection{Correlation between MSP mass and orbital period?}
Progenitors of NSs in wide orbits are less stripped prior to their
explosion compared to those in close systems \citep{ywl10}.  Hence, it
is possible that the resulting NS masses could be somewhat larger in
wider systems. To probe such a relation is difficult because it is
masked by the subsequent LMXB accretion phase, which recycles the NS to
become an MSP.  Based on binary stellar evolution
theory, \citet{ts99a} argued for an anti-correlation between the
amount of mass accreted by the NS, $\Delta M_{\rm NS}$ and orbital
period.  Despite this effect, however, we still find evidence for
slightly more massive MSPs in wider systems.  Using the sample of MSPs
with He~white dwarfs studied in Section\,5.2, we find that the overall average
NS mass is larger by $0.06\pm0.01$\;M$_{\odot}$ in systems with
$P_{\rm b}>30\;{\rm days}$, compared to systems with $P_{\rm
  b}<30\;{\rm days}$.  Hence, we would expect the mass difference to
be even larger at their birth, thereby supporting the evidence for the
hypothesis that wide binaries, in general, produce more massive NSs at
birth.  Larger MSP masses in wider orbits can also help explain the
results of \cite{sfl+05}, who found it difficult to reconcile the
$M_{\rm WD}-P_{\rm b}$~relation with observational data based on a
statistical analysis assuming the same MSP mass in all binaries.

\subsection{Constraints on the Maximum Mass}
High-mass NSs place stringent constraints on the EoS beyond the
nuclear saturation density. The most massive known NS with a precisely
measured mass is PSR\,J0348+0432 with $M=2.01(4)$\,M$_{\odot}$, which
is commonly adopted at face value as the limit for the maximum NS
mass. Here, we demonstrate an alternative method which relies on
Bayesian inference of the MSP mass distribution properties.

\cite{kkd+13}, who took a similar approach, identified the maximum NS
mass with the tail of their inferred MSP mass distribution. In
reality, this limit has very little physical relevance. This is
because the simple empirical models used to fit the observed masses
are not likely to be true representations of the underlying mass distribution and any
statistical model can have an unaccounted for mass cut-off at the high
mass end. Consequently, extrapolation to high masses is of limited
value, as there is no guarantee that the observed masses carry
information for the true maximum NS mass. Even if this were the case,
the $3\sigma$ limit of the mass distribution, adopted by \cite{kkd+13}
does not prevent the existence of NSs with larger masses.

Here, we have made use of the fact that the EoS can introduce a
high-mass cut-off in the NS distribution.  A search for a truncation
in the currently observed masses with our Bayesian inference method,
yields a limit for the maximum mass of $M_{\rm max} \ge
2.018$\,M$_{\odot}$ at 98\% C.L or 1.923\,M$_{\odot}$ at 99.98\% C.L.
Interestingly, the former appears to be insensitive to our choice for
the model distribution.  This method yields a more robust constraint,
in the sense that it is derived from the likelihoods of all massive
NSs.

\subsection{Accelerating the discovery of massive NSs and 
prospects for measuring the maximum NS mass} The Bayesian framework
used in our study also allows us to estimate the number of mass
measurements required for a precision localization of a high-mass
truncation in the underlying mass distribution.  Our estimates suggest
that if the maximum mass is smaller than $\sim 2.15$\,M$_{\odot}$,
then the measurement of 350 MSP masses following the inferred
distribution suffices to localize $M_{\rm max}$ with a precision of
$\sim 5\%$.  This number of inferred MSP masses should be possible
with the upcoming SKA surveys.

Obviously, the most important factor impacting the detection of
M$_{\rm max}$ is its actual value. Constraints on the NS radius from
bursting and quiescent LMXB sources currently favor softer EoSs which
cannot support NSs with masses much greater than
2\,M$_{\odot}$ \citep{of16}. Hence, it is possible that stringent constraints on
M$_{\rm max}$ can be achieved sooner. Improving overall on measurement
uncertainties, e.g., by increasing the observing cadence may help as
well.

Another possible strategy would be to focus only on those NSs
occupying the high-mass tail of the distribution.  In Section\,5.2, we
argued that, for MSPs with white dwarf companions, it is possible to identify
potential high-mass candidates by making use of the $M_{\rm WD}-P_{\rm
  b}$ correlation for post-LMXB systems. Of the  
binaries shown in Table\,5, some 20 have high probability ($\gtrsim 80\%$)
for having a mass above 1.8\,M$_{\odot}$.  Finally, a complementary
approach would be to focus on special types of systems such as
eclipsing MSPs and DNSs. For the former, existing mass constraints
could be improved by exploring alternative methods, such as high
resolution wide-band spectroscopy, or scintillometry
\citep{pmd+14}. DNSs such as the double pulsar \citep{ksm+06} on the
other hand, may make it possible to identify ``special'' NSs, such as
those formed through an electron-capture SN \citep{plp+04} or an
ultra-stripped iron core-collapse SN \citep{tlm+13,tlp15}, which have
the potential to place direct constraints on the NS gravitational
binding energy and consequently on the EoS.  In addition, for the
double pulsar, the measurement of the Lense-Thirring precession may
soon result in the first measurement of a NS moment of inertia
\citep{kw09,kwkl16}, which results in direct constraints on the NS radius and the EoS \citep{rop16}.

\acknowledgements We thank Norbert Wex and Michael Kramer for discussions and comments on the manuscript. 
JA, TMT \& F\"O wish to thank  the Munich Institute for Astro and Particle Physics
(MIAPP) of the DFG cluster of excellence ``Origin and Structure of the
Universe'' for their support and hospitality  during the workshop "The many faces of neutron stars" which inspired this work. 
JA is a Dunlap Fellow at the Dunlap Institute for
Astronomy and Astrophysics at the University of Toronto.  The Dunlap
Institute is funded by an endowment established by the David Dunlap
family and the University of Toronto.  
TMT acknowledges the receipt of DFG Grant TA\,964/1-1. 
F\"O acknowledges support from NSF grant AST 1108753 and
NASA grant NNX16AC56G. PF acknowledges financial support by the European Research Council for the ERC Starting grant AST 1108753. 
We have made extensive use of NASA's Astrophysics Data System.

\appendix

\section{The mass of PSR\,J1012+5307}
PSR\,J1012+5307 is a 5.3\,ms pulsar with a low-mass He-white dwarf companion in
a 14.4\,h orbit. A spectroscopic analysis of the white dwarf was performed by
two groups \citep{vbk96,cgk98}, who found different values for both
the atmospheric properties and radial velocities. A subsequent
analysis of the results revealed that the differences in the
velocities were caused by a bias in the \cite{vbk96} analysis. A
reanalysis of the high S/N Keck data from \cite{vbk96} yields $K_{\rm
  WD} = 199\pm10$\,km\,s$^{-1}$ for the semi-amplitude of the white dwarf's
orbital radial velocity, and $q\equiv m_{\rm p}/m_{\rm c} =
10.0\pm0.7$ for the mass ratio, in agreement with \cite{cgk98}. The
discrepancy on the value of the surface gravity was traced to slight
differences in the input physics of the atmospheric models used by the
two teams \citep{vbjj05}.

 Interestingly, both studies find the same mass for the pulsar,
 $m_{\rm p} = 1.6(2)$\,M$_{\odot}$, but again due to the different white dwarf 
 mass-radius relations adopted in their analysis.  These models did
 not properly account for finite-temperature corrections nor the
 effect of an extended hydrogen envelope. A follow-up study by
 \cite{dsbh98}, based on the (biased) values of \cite{vbk96}, using
 appropriate input physics, found $m_{\rm c} = 0.19(2)$\,M$_{\odot}$
 implying $m_{\rm p} = 1.9(3)$\,M$_{\odot}$.

Another important effect that became evident after these early He-white dwarf 
studies, is a bias in 1-D atmospheric models for relatively cool white dwarfs
\citep{tls+11}. Recent work demonstrates that this effect is caused by
the imperfect scheme used to model convective transport in 1D models,
with full corrections based on 3D DA atmospheres now available for the
entire parameter space relevant to MSP companions
\citep{tls+13,tgk+15}.  Finally, the recent detection of pulsational
instabilities in white dwarfs with similar temperatures and masses, allows us
to place further constraints on the mass of the system: the surface
gravity reported by \cite{cgk98} would place PSR\,J1012+5307 in the
middle of the instability strip, as derived empirically by
\cite{gkb+15}. Such pulsations are not detected \citep{khg+15},
implying that the true (1D) atmospheric parameters must be close to
those reported by \cite{vbk96}.

To derive the mass estimate reported in Table\,1, we start with a
simulated distribution of atmospheric parameters with $T_{\rm eff}=
8550(50)$\,K and $\log_{10}g = 6.75(1)$\,dex, following \cite{vbk96},
but with slightly increased error estimates, to account for possible
remaining uncertainties, and because we do not have access to the full
covariance matrix of their atmospheric fit. We then map these samples
to 3D-corrected values, using the relations of \cite{tgk+15} , and
then to a mass-radius distribution using the models of \cite{amc13},
which have been shown to yield reliable parameters for similar He-white dwarfs
\citep[e.g.,][and references therein]{aks+16}. Finally, we derive the mass of the pulsar,
$m_{\rm p} = 1.83(11)$\,M$_{\odot}$ using the mass-ratio estimate
discussed above. A follow-up spectroscopic study of PSR\,J1012+5307 to
verify this estimate is in progress (Gemini project: GN-2016A-Q-70).

\section{Mass constraints for MSPs with He-white dwarf companions}
Table\,5 shows the predictions for the masses of MSPs with He~white dwarf
companions described in Section\,5.2. The companion masses, orbital
periods, dispersion measures and inferred distances based on the
NE\,2001 \citep{ne2001} model for the distribution of free electrons
in the Galaxy are also shown. Finally, we also calculate the pulsar
mass for fixed inclination angles ($i =  30^{\rm o}$ and 
$60^{\rm o}$), and the inclination corresponding to a ``canonical''
pulsar mass of 1.4\,M$_{\odot}$. The last column shows the likelihood for the pulsar to have a mass above 1.8\,M$_{\odot}$. 

\centering

\begin{longtable}{lccccccccc}

\hline
Name & P$_{0}$ & DM & $d$ & P$_{\rm b}$ & m$_{\rm c}^{\rm ts99}$ & m$_{\rm p}^{30^{\rm o}}$  &  m$_{\rm p}^{60^{\rm o}}$ &  $i(m_{\rm p} = 1.4$\,M$_{\odot}$) & $\mathscr{L}(m_{\rm p} > 1.8$\,M$_{\odot}$) \\
 & (s) & (cm$^{-3}$\,pc) & $(kpc)$ & (days) & M$_{\odot}$ &  M$_{\odot}$ & M$_{\odot}$ & $^{\rm o}$ & \\
\hline 
PSR\,J0034-0534 & 0.0019 & 13.77 & 0.98 & 1.59 & 0.21 & 0.75  & 1.98 & 44.82 & 0.73 \\
PSR\,J0101-6422 & 0.0026 & 11.93 & 0.73 & 1.79 & 0.21 & 0.64  & 1.74 & 49.74 & 0.66 \\
PSR\,J0218+4232 & 0.0023 & 61.25 & 3.15 & 2.03 & 0.22 & 0.57  & 1.57 & 54.00 & 0.55 \\
PSR\,J0437-4715 & 0.0058 & 2.64 & 0.16 & 5.74 & 0.24 & 0.94  & 2.45 & 38.46 & 0.62 \\
PSR\,J0557+1550 & 0.0026 & 102.57 & 5.65 & 4.85 & 0.24 & 0.50  & 1.44 & 58.50 & 0.23 \\
PSR\,J0613-0200 & 0.0031 & 38.78 & 0.90 & 1.20 & 0.21 & 0.85  & 2.20 & 41.43 & 0.65 \\
PSR\,J0614-3329 & 0.0031 & 37.05 & 2.96 & 53.58 & 0.32 & 0.39  & 1.29 & 64.65 & 0.00 \\
PSR\,J1017-7156 & 0.0023 & 94.22 & 0.26 & 6.51 & 0.24 & 0.55  & 1.57 & 54.11 & 0.51 \\
PSR\,J1045-4509 & 0.0075 & 58.17 & 0.23 & 4.08 & 0.23 & 0.71  & 1.91 & 46.27 & 0.74 \\
PSR\,J1056-7117 & 0.0263 & 93.04 & 5.27 & 9.14 & 0.25 & 1.23  & 3.14 & 32.43 & 0.59 \\
PSR\,J1125-5825 & 0.0031 & 124.79 & 2.98 & 76.40 & 0.33 & 0.47  & 1.51 & 56.35 & 0.46 \\
PSR\,J1216-6410 & 0.0035 & 47.40 & 1.71 & 4.04 & 0.23 & 0.73  & 1.97 & 45.25 & 0.75 \\
PSR\,J1231-1411 & 0.0037 & 8.09 & 0.45 & 1.86 & 0.21 & 0.47  & 1.34 & 62.78 & 0.00 \\
PSR\,J1232-6501 & 0.0883 & 239.40 & 10.00 & 1.86 & 0.21 & 0.76  & 2.00 & 44.57 & 0.71 \\
PSR\,J1327-0755 & 0.0027 & 27.91 & 2.17 & 8.44 & 0.25 & 0.42  & 1.27 & 65.99 & 0.00 \\
PSR\,J1405-4656 & 0.0076 & 13.88 & 0.74 & 8.96 & 0.25 & 0.48  & 1.41 & 59.58 & 0.12 \\
PSR\,J1431-5740 & 0.0041 & 131.46 & 4.07 & 2.73 & 0.22 & 0.68  & 1.83 & 47.71 & 0.71 \\
PSR\,J1455-3330 & 0.0080 & 13.57 & 0.74 & 76.17 & 0.33 & 0.52  & 1.61 & 53.39 & 0.58 \\
PSR\,J1543-5149 & 0.0021 & 50.93 & 1.46 & 8.06 & 0.25 & 0.41  & 1.25 & 67.32 & 0.00 \\
PSR\,J1545-4550 & 0.0036 & 68.39 & 2.01 & 6.20 & 0.24 & 0.82  & 2.17 & 42.05 & 0.65 \\
PSR\,J1600-3053 & 0.0036 & 52.33 & 2.40 & 14.35 & 0.27 & 0.55  & 1.60 & 53.45 & 0.54 \\
PSR\,J1622-6617 & 0.0236 & 88.02 & 4.66 & 1.64 & 0.21 & 1.56  & 3.83 & 27.96 & 0.58 \\
PSR\,J1640+2224 & 0.0032 & 18.43 & 1.19 & 175.46 & 0.37 & 0.67  & 2.01 & 45.27 & 0.68 \\
PSR\,J1643-1224 & 0.0046 & 62.41 & 0.42 & 147.02 & 0.36 & 2.40  & 5.94 & 21.73 & 0.56 \\
PSR\,J1708-3506 & 0.0045 & 146.73 & 3.50 & 149.13 & 0.36 & 1.45  & 3.78 & 29.37 & 0.57 \\
PSR\,J1709+2313 & 0.0046 & 25.35 & 1.83 & 22.71 & 0.28 & 0.33  & 1.12 & 77.73 & 0.00 \\
PSR\,J1711-4322 & 0.1026 & 191.50 & 4.17 & 922.47 & 0.48 & 1.53  & 4.10 & 28.58 & 0.00 \\
PSR\,J1713+0747 & 0.0046 & 15.97 & 1.05 & 67.83 & 0.33 & 0.41  & 1.36 & 61.51 & 0.02 \\
PSR\,J1732-5049 & 0.0053 & 56.82 & 1.81 & 5.26 & 0.24 & 0.59  & 1.65 & 51.85 & 0.59 \\
PSR\,J1745-0952 & 0.0194 & 64.47 & 2.38 & 4.94 & 0.24 & 1.44  & 3.58 & 29.52 & 0.58 \\
PSR\,J1751-2857 & 0.0039 & 42.81 & 1.44 & 110.75 & 0.35 & 0.98  & 2.68 & 36.93 & 0.60 \\
PSR\,J1801-3210 & 0.0075 & 177.71 & 5.08 & 20.77 & 0.28 & 1.24  & 3.18 & 32.35 & 0.59 \\
PSR\,J1803-2712 & 0.3344 & 165.50 & 3.62 & 406.78 & 0.42 & 2.24  & 5.65 & 22.84 & 0.00 \\
PSR\,J1804-2717 & 0.0093 & 24.67 & 1.17 & 11.13 & 0.26 & 0.55  & 1.58 & 53.97 & 0.52 \\
PSR\,J1811-2405 & 0.0027 & 60.60 & 1.70 & 6.27 & 0.24 & 0.35  & 1.11 & 80.04 & 0.00 \\
PSR\,J1813-2621 & 0.0044 & 112.52 & 3.37 & 8.16 & 0.25 & 0.58  & 1.65 & 52.08 & 0.58 \\
PSR\,J1825-0319 & 0.0046 & 119.50 & 3.26 & 52.63 & 0.31 & 0.97  & 2.62 & 37.28 & 0.60 \\
PSR\,J1835-0114 & 0.0051 & 98.00 & 2.67 & 6.69 & 0.24 & 0.63  & 1.74 & 49.86 & 0.66 \\
PSR\,J1841+0130 & 0.0298 & 125.88 & 3.19 & 10.47 & 0.26 & 1.99  & 4.87 & 24.07 & 0.57 \\
PSR\,J1844+0115 & 0.0042 & 148.22 & 3.45 & 50.65 & 0.31 & 1.48  & 3.78 & 28.97 & 0.57 \\
PSR\,J1850+0124 & 0.0036 & 118.89 & 2.97 & 84.95 & 0.34 & 0.57  & 1.72 & 50.72 & 0.65 \\
PSR\,J1853+1303 & 0.0041 & 30.57 & 1.60 & 115.65 & 0.35 & 0.65  & 1.92 & 46.70 & 0.74 \\
PSR\,J1857+0943 & 0.0054 & 13.30 & 0.90 & 12.33 & 0.26 & 0.38  & 1.19 & 71.33 & 0.00 \\
PSR\,J1901+0300 & 0.0078 & 253.89 & 5.50 & 2.40 & 0.22 & 0.84  & 2.21 & 41.43 & 0.65 \\
PSR\,J1904+0412 & 0.0711 & 185.90 & 4.01 & 14.93 & 0.27 & 0.48  & 1.44 & 58.49 & 0.28 \\
PSR\,J1910+1256 & 0.0050 & 38.07 & 1.95 & 58.47 & 0.32 & 0.85  & 2.35 & 40.18 & 0.62 \\
PSR\,J1911-1114 & 0.0036 & 30.98 & 1.59 & 2.72 & 0.22 & 1.09  & 2.77 & 35.18 & 0.60 \\
PSR\,J1918-0642 & 0.0076 & 26.55 & 1.40 & 10.91 & 0.26 & 0.38  & 1.20 & 70.32 & 0.00 \\
PSR\,J1935+1726 & 0.0042 & 61.60 & 3.11 & 90.76 & 0.34 & 0.73  & 2.10 & 43.70 & 0.66 \\
PSR\,J1955+2908 & 0.0061 & 104.50 & 5.39 & 117.35 & 0.35 & 1.15  & 3.07 & 33.66 & 0.59 \\
PSR\,J2016+1948 & 0.0649 & 33.81 & 1.83 & 635.02 & 0.45 & 0.68  & 2.12 & 44.11 & 0.00 \\
PSR\,J2017+0603 & 0.0029 & 23.92 & 1.32 & 2.20 & 0.22 & 0.52  & 1.47 & 57.28 & 0.35 \\
PSR\,J2019+2425 & 0.0039 & 17.20 & 0.91 & 76.51 & 0.33 & 0.32  & 1.16 & 73.39 & 0.00 \\
PSR\,J2033+1734 & 0.0059 & 25.08 & 1.37 & 56.31 & 0.32 & 0.88  & 2.42 & 39.36 & 0.62 \\
PSR\,J2043+1711 & 0.0024 & 20.71 & 1.13 & 1.48 & 0.21 & 0.53  & 1.48 & 57.05 & 0.34 \\
PSR\,J2129-5721 & 0.0037 & 31.85 & 0.40 & 6.63 & 0.24 & 1.07  & 2.76 & 35.41 & 0.60 \\
PSR\,J2229+2643 & 0.0030 & 23.02 & 1.43 & 93.02 & 0.34 & 2.08  & 5.19 & 23.63 & 0.56 \\
PSR\,J2236-5527 & 0.0069 & 20.00 & 2.03 & 12.69 & 0.26 & 0.45  & 1.36 & 61.74 & 0.00 \\
PSR\,J2317+1439 & 0.0034 & 21.91 & 1.89 & 2.46 & 0.22 & 0.56  & 1.55 & 54.64 & 0.48 \\
\hline
\caption{Predictions for MSPs with He white-dwarf companions}

\end{longtable}

\bibliography{author}

\end{document}